\documentclass[aps,prb,amsmath,twocolumn]{revtex4}
\usepackage{graphicx,dcolumn,bm,amssymb,amsmath}
\usepackage{color,wrapfig,braket}
\usepackage{hyperref,placeins,ulem}
\usepackage[caption=false]{subfig}

\usepackage{epsfig}
\usepackage{verbatim}
\usepackage{yhmath}

\begin{document}

\title{Layer construction of 3D topological states and string braiding statistics}
\author{Chao-Ming Jian and Xiao-Liang Qi}
\affiliation{Department of Physics, Stanford University, Stanford, California 94305, USA.}
\date{\today}

\begin{abstract}
While the topological order in two dimensions has been studied extensively since the discover of the integer and fractional quantum Hall systems, topological states in 3 spatial dimensions are much less understood. In this paper, we propose a general formalism for constructing a large class of three-dimensional topological states by stacking layers of 2D topological states and introducing coupling between them. Using this construction, different types of topological states can be obtained, including those with only surface topological order and no bulk topological quasiparticles, and those with topological order both in the bulk and at the surface. For both classes of states we study its generic properties and present several explicit examples. As an interesting consequence of this construction, we obtain example systems with nontrivial braiding statistics between string excitations. In addition to studying the string-string braiding in the example system, we propose a generic topological field theory description which can capture both string-particle and string-string braiding statistics.
Lastly, we provide a proof of a general identity for Abelian string statistics, and discuss an example system with non-Abelian strings.
\end{abstract}
\maketitle

\tableofcontents

\section{Introduction}
\label{intro}
Some of the most important discoveries in condensed matter physics over the last few decades have been about topological states of matter. Topological states are gapped states of matter that are not characterized by broken symmetries and local order parameters, but are still distinct phases from a trivial insulator. A subclass of topological states of matter is the topologically ordered states, which are stable against any local perturbations and have topologically protected properties including fractional quasi-particle statistics, ground state degeneracy determined by topology of the spatial manifold, etc\cite{Wen2004}. Since the discovery of integer and fractional quantum Hall effects\cite{klitzing1980,tsui1982}, 2D topologically ordered states have been characterized and constructed by many different approaches, such as ideal wavefunctions\cite{laughlin1983,Moore1991,wen2008a,bernevig2008}, topological field theories\cite{wen1992,zhang1992}, exact lattice models\cite{Kitaev2003,wen2003,kitaev2006,LevinWen} and tensor networks\cite{Verstraete2006,gu2009,zaletel2012,estienne2013,wahl2013,dubail2013}.
The general structure of topological states in 2D is governed by the mathematical frame work of tensor category theory\cite{kitaev2006,zhwang2010,nayak2008}. 
Since the discovery of topological insulators\cite{hasan2010,qiRMP2011}, new types of topologically ordered states with nontrivial symmetry properties have also been studied, known as symmetry protected topological (SPT) states and symmetry enriched topological (SET) states.\cite{chen2011b,hung2013,vishwanath2013,senthil2014}

In contrast to the progress in 2D, much less is known about topologically ordered states in 3D. Since there is not yet a general framework for 3D topologically ordered states, building different models with topological order is an important method that can help us understand the general features of 3D topological order. The 3D topological insulators\cite{hasan2010,qiRMP2011} can be generalized to fractional topological insulators with topological order and fractionally charged partons\cite{bernevig2006a,levin2009,maciejko2010b,swingle2011b}. Some exact solvable models in 2D can be generalized to 3D. For example, the 3D generalization of the toric code model can capture the topological order of 3D lattice discrete gauge theory in which the ground state degeneracy is associated to the non-trivial $1$-cycles and the point particles have non-trivial mutual braiding statistics with flux string excitations. The Levin-Wen models can be generalized to the Walker-Wang models\cite{walker2012,simon2013} which describe 3D gapped states with interesting surface topological order and 3D bulk topological order that generalizes the 3D lattice gauge theory. In 3D, the statistics between particles is a representation of the permutation group instead of braiding group in 2D, so that there cannot be fractional statistics for point-like particles\cite{doplicher1971,doplicher1974} (although non-trivial statistics have been shown to be possible for point-like monopole defects\cite{teo2010,ran2011,freedman2011,freedman2011b}). However, 3D topologically ordered states can have 1D string-like excitations in addtion to particles. The existence of string excitation enables rich possibilities of braiding statistics. Recently, the topological properties of the lattice models of the 3+1D Dijkgraaf-Witten lattice gauge theory\cite{dijkgraaf1990,Levin2014StrBrd,Ran2014StrBrd,JuvenWenStrBrd2014} and 3D generalization of quantum double models\cite{Shor2011,Wen2014StrBrd} are studied. In these models, string excitations have non-trivial fusion and braiding statistics.

In this paper, we propose a different approach of constructing 3D topological states named as the layer construction. In this construction, 3D topological states can emerge in a system with stacked layers of 2D topological states and properly engineered inter-layer couplings. This type of construction is first proposed in an special example of 3D boson topological insulators in Ref. \onlinecite{wang2013}. It is reminiscent of the wire construction of 2D topological states\cite{sondhi2001,kane2002,teo2014}. It is shown that if we take an array of the 1D wires of Luttinger liquids and engineer the inter-wire coupling, this array of wires can effectively form a 2D quantum Hall liquid. Depending on the coupling, a large class of fractional quantum Hall states can be constructed. Motivated by the success of the wire construction for 2D topological states and by the example in Ref. \onlinecite{wang2013}, we study in this paper the layer construction of 3D states as a more general formalism of constructing and characterizing 3D topological states. In this general formalism, we consider coupling layers of 2D Abelian topological states where the consequence of the coupling is to induce anyon condensation of some composite particle formed by topological quasiparticles in two adjacent layers. By several explicit examples, we show that this approach can describe a large class of different 3D topological states, including those with only surface 2D topological order, and those with bulk and/or surface topological order. 
For the systems with only surface topological order and no bulk topological order, we construct a series of examples which are 3D gapped states with interesting chiral 2D topological order on their surfaces, although each layer of 2D states in the construction is non-chiral. 
This is similar to the observation of Ref. \onlinecite{simon2013,XieChen2013} in Walker-Wang models.  We obtain general criteria for the absence of 3D topological order and the characterization of surface topological order. With these criteria satisfied, we show how to determine the surface topological order on top, bottom and side surfaces in a general situation and show the equivalence of different surfaces, which proves that the states obtained in this construction are topologically isotropic. For more general states with nontrivial 3D topological order, our layer construction enables description of the bulk point-particles and string-like excitations. We provide several examples of different topological orders, including a ``convectional" 3D topological order that resembles the lattice gauge theory, and more general systems with coexisting bulk and surface topological order. As a particularly interesting feature of this construction, we can construct some 3D topologically ordered states with non-trivial string-string braiding statistics, in addition to the usual particle-string braiding statistics. We provide a field theory description of the 3D topological order and the string-string and particle-string braiding statistics. We also discuss the general features of string braiding, and provide a more general proof of a constraint on string braiding statistics proposed recently by Ref.\cite{Levin2014StrBrd}. In the end,
give an example system with non-Abelian string braiding statistics, to show that more general topological orders are possible in 3D. 

The rest of the paper is organized as follows. In Sec. \ref{generalconstrtuction}, we propose a general framework for the layer construction with layers of 2D Abelian topological states. We introduce the coupling between layers by turning on anyon condensations\cite{maissam2013ac1,maissam2013ac2,levin2013ac,Kong2013} for certain composite quasi-particles living in two adjacent layers. We provide general 
requirements on such condensed particles. In Sec. \ref{pSTO}, we focus on the construction of 3D gapped states with only 2D topological order on their surfaces and no bulk topological order. Sec. \ref{EGpSTO} presents several examples, and Sec. \ref{GCpSTO} discusses the general criteria of surface-only topological order. Sec. \ref{surfaceOrder} studies the surface topological order on top, bottom and side surfaces and prove their equivalence.  
In Sec. \ref{3DTO}, we discuss more general states with bulk topological order. Sec. \ref{Convent3DTO} discusses the conventional topological order of the type of discrete gauge theories. Sec. \ref{Coexist} discusses examples with bulk topological order coexisting with surface topological order. Sec. \ref{GenericState} discusses the most nontrivial types of system with nontrivial string-string braiding statistics.
In Sec. \ref{FieldTheory}, we provide a field theory description of the topological order and the string braiding statistics for the layer constructed models. Sec. \ref{GeneralStrBrd} is devoted to more general discussions on string braiding statistics. In Sec. \ref{GeneralA} we prove a general identity that should be satisfied by Abelian string braiding statistics. In Sec. \ref{GeneralB} we 
we provide an example of non-Abelian strings and their statistics. The final section \ref{conclusion} concludes the paper with some discussions on open questions and future directions.


\section{General Setting of the Layer Construction}
\label{generalconstrtuction}
In this section, we will introduce the general formalism for the construction of $3$D topological states from layers of $2$D topological states. In general, one can start with any topological state in each $2$D layer. For concreteness of the discussion, we restrict ourselves to only consider layers of $2$D Abelian topological states. Before we introduce the general framework of the layer construction, we will first briefly review the theory of $2$D Abelian topological states and the condensation of certain quasi-particles in these states. $2$D Abelian topological states can be described by the Abelian Chern-Simons theory with a K-matrix, of which the Lagrangian\cite{wen1992kMatrix} is given by:
\begin{align}
L_{CS}=\int dx dy dt \frac{1}{4\pi}\epsilon^{\mu\nu\lambda} K_{IJ} a^I_\mu \partial_\nu a^J_\lambda,
\end{align}
where $a^I$ for $I=1,...,r$ are compact $U(1)$ gauge fields and $K$ is a non-singular integer symmetric matrix. Here We've denoted the dimension of the K-matrix as $r=\text{dim}(K)$. The quasi-particles in this theory are labelled by $r$-component integer-valued vectors. The topological spin of the quasi-particle labelled by the integer vector $l$ is given by $\theta_l=\pi l^T K^{-1} l$, and the mutual statistics of two quasiparticles $l$ and $l'$ is $\theta_{ll'}=2\pi l^T K^{-1} l'$. A particle labeled by $l=K v$, with $v$ an $r$-component integer vector, is considered as a local particle in the theory. In this theory, we can further consider the condensation of a subgroup of the quasiparticles $M_{\text{Lag}}$, called a ``Lagrangian subgroup", which has the following properties\cite{levin2013ac,maissam2013ac1,maissam2013ac2}:
\begin{enumerate}
  \item $e^{i\theta_m}=e^{i\theta_{mm'}}=1$, for all $m,m'\in M_{\text{Lag}}$;
  \item For all $l\notin M_{\text{Lag}}$, $e^{i\theta_{lm}}\neq 1$ for at least one $m\in M_{\text{Lag}}$.
\end{enumerate}
The condensation of a Lagrangian subgroup drives a transition from the Abelian topological state to a topologically trivial phase. For each Lagrangian subgroup, there is an equivalent description of the condensate using another set of particles $M_\text{null}$\cite{levin2013ac,maissam2013ac1,maissam2013ac2}. 
We refer to the particles in this set as null particles. This set $M_\text{null}$ is closed under particle fusion and satisfies the following conditions:
\begin{enumerate}
  \item $\theta_m=\theta_{mm'}=0$, for all $m,m'\in M_{\text{null}}$;
  \item $M_\text{null}$ is generated by $\text{rank}(M_\text{null})=r/2$ null particles $m_i$, $i=1,...,r/2$.
\end{enumerate}
Notice that for Lagrangian subgroup or the null particle set to exist, the dimension of the K-matrix $r$ has to be even. It has been proven that the Lagrangian subgroups $M_{\text{Lag}}$ and the null particle sets $M_{\text{null}}$ are in one-to-one correspondence to each other. \footnote{More rigorously, this correspondence only applies when one is allowed to topologically deform the system by adding topologically trivial blocks to the $K$ matrix.
More details were discussed in Ref. \onlinecite{levin2013ac,maissam2013ac1,maissam2013ac2}. } 
The condensation of the particles of the null set $M_{\text{null}}$ will induce the condensation of all the particles in its corresponding Lagrangian subgroup $M_{\text{Lag}}$ modulo the local particles that can be always thought of as condensed, and vice versa.

\begin{figure}[tb]
\centerline{
\includegraphics[width=3
in]{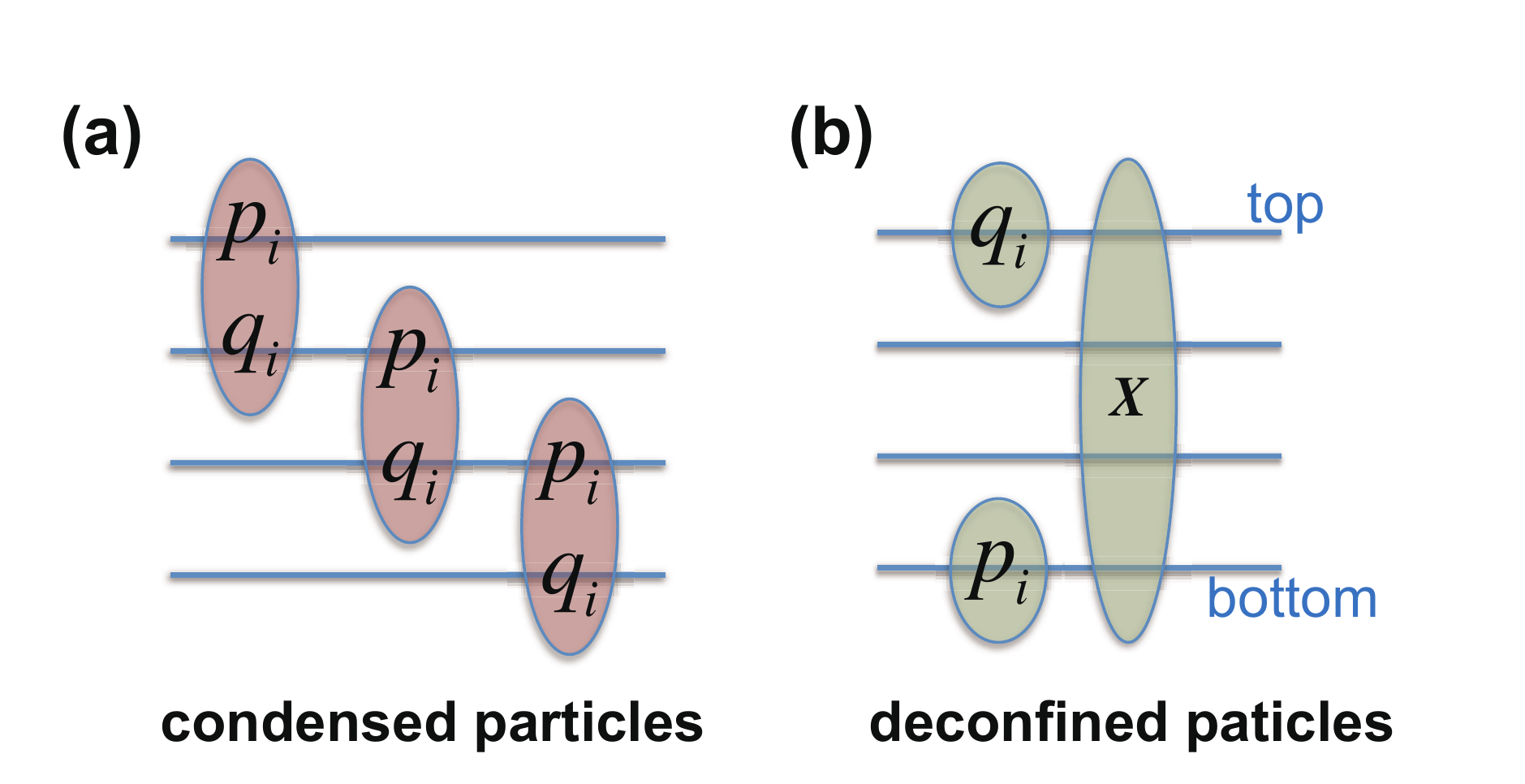}
}
\caption{\label{nullparticle} (a) Each line represents a layer of 2D Abelian topological state, and each brown circle represents a composite particle, the condensation of which will introduce the coupling between layers. Here each composite particle lives only in two consecutive layers, and the component of the particle in upper and lower layers is given by $p_i$'s and $q_i$, respectively. (b) For open boundary system in the $z$ direction, $q_i$'s ($p_i$'s) are deconfined particles on the top (bottom) surface which will form surface topological order. Depending on the choice of $\{p_i\}$ and $\{q_i\}$, there can be deconfined particles (represented by ``$X$") in the 3D bulk that organize themselves to form 3D topological orders.
}
\end{figure}

Now we can start introducing the general setting of the layer construction of the $3$D topological states. First, we consider stacked layers of the identical $2$D Abelian topological states with the K-matrix $K$ for each layer. Before we introduce the coupling between the layers, we can view the system as a $2$D state using the $2+1$D Chern-Simons theory with an extended K-matrix $\mathcal{K}$ given by
\begin{align}
\mathcal{K}=K \otimes \left(
\begin{array}{ccc}
1 & & \\ & 1 & \\ & & \ddots
\end{array}
\right)_{L\times L},
\end{align}
where $L$ stands for the number of layers in the system. We can turn on the ``local" coupling between the layers by introducing the condensation of composite quasi-particles that lives in finite range of layers. Without the loss of generality, we only consider the composite particles that live in two consecutive layers as is shown in Fig. \ref{nullparticle} (a). These particles, which we label as $n_i^{(m)}$, take the form of
\begin{align}
n_i^{(m)}=p_i \otimes e_m+q_i \otimes e_{m+1},  \label{nullparticles}
\end{align}
where $p_i$'s and $q_i$'s are $r$-component integer vectors that labels the quasi-particle type in one single layer and $e_m$ is a $L$ component unit vector with the $m^\text{th}$ entry $1$ and all the rest $0$. The superscript $m$ is effectively the layer index, which will later be identified as the ``lattice coordinate" in the $z$ direction. For a system that is periodic in the perpendicular $z$ direction, we need to consider $m=1,2,...,L$ with the identification that $e_{L+1}=e_1$ while, for the open boundary condition in the $z$ direction, we only need to consider the condensation of null particles with $m=1,2,...,L-1$. The subscript $i$ labels the different types of the condensed quasi-particles $n_i^{(m)}$. It should be noticed that the condensed particles are completely determined by the sets $\{p_i\}$ and $\{q_i\}$ that are independent from the layer index $m$. Thus, the condensate of these particles is invariant under the cyclic permutation between the layer, namely under $m\rightarrow m+1$, and therefore is translationally invariant in the $z$ direction. For the sets $\{p_i\}$ and $\{q_i\}$, we require them to satisfy the following null conditions:
\begin{align}
& p^T_i K^{-1} p_j+ q^T_i K^{-1} q_j=0, \nonumber \\
& p^T_i K^{-1} q_j=0,~~~ \forall~i \text{ and }j,
\label{nullcondition}
\end{align}
such that the mutual statistics and the topological spins of the condensed particles $n_i^{(m)}$ is guaranteed to be trivial:
\begin{align}
n_i^{(m)T}\mathcal{K}^{-1} n_{i'}^{(m')}=0,~~~~\forall i,i',m,m'.
\end{align}
This condition is similar to the null conditions of the null particle set $M_\text{null}$ for $2$D Abelian topological state introduced above. Moreover, in analogy to the requirement to the rank of the null set $\text{rank}(M_\text{null})$, here we also require that
\begin{align}
 \text{number of } p_i\text{'s} =  \text{number of } q_i\text{'s} = \text{dim}(K)/2=r/2.  \label{nullcounting}
\end{align}
That is to say the index $i$ takes value $i=1,2,...,r/2$. In this counting requirement, we have implicitly assume that for a given $m$, the choice of $\{p_i\}$ and $\{q_i\}$ guarantees that all particles $n_i^{{m}}$ for $i=1,...,r/2$ are linearly independent. Again, we only consider cases with even $r$. Here the choice of the sets $\{p_i\}$ and $\{q_i\}$ is subject to an equivalence relation. For any integer matrix $W$ with $\det(W)=1$, we can define new sets of $\{p'_i\}$ and $\{q'_i\}$ by the linear transformation $p'_j=\sum_i W_{ij} p_i$ and $q'_j=\sum_i W_{ij} q_i$ so that the condensed particle $\{n'^{(m)}_{i}\}$ defined by the new sets satisfy $n'^{(m)}_j=\sum_i W_{ij} n^{(m)}_i$. It is obvious that the condensation of $\{n'^{(m)}_{i}\}$ is equivalent to that of $\{n^{(m)}_{i}\}$. Therefore, the choice of sets $\{p_i\}$ and $\{q_i\}$ is defined modular a linear transformation by 
an integer uni-determinat matrix $W$.


To summarize the general frame work of this construction, we start with layers of 2D Abelian topological states with each layer described by a K-matrix theory. Then, in order to couple the layer together to form a 3D topological state, we turn on the condensation of the particles of the form given in Eq. (\ref{nullparticles}) which is determined by the sets $\{p_i\}$ and $\{q_i\}$ that satisfy the null conditions in Eq. (\ref{nullcondition}) and the counting requirements in Eq. (\ref{nullcounting}). 
Before we move on to more detailed analysis of the layer construction, a couple of comments are in order. Firstly, we argue that the layer construction indeed generate a genuine 3D state. The reasons for this is the following: 1) It is shown in Ref. \onlinecite{levin2013ac,maissam2013ac1, maissam2013ac2} that the condensation of the null particles can be induced by introducing coupling using only local variables. Here the locality is understood in the 2D sense; 2) We only condense composite particles that live in two consecutive layers. Therefore, the locality in the $z$ direction is also guaranteed; 3) The condensation of each quasi-particle is a phase transistion from one gapped state to another gapped state. Therefore, the resulting states of the layer construction should also be a gapped 3D state; 4) The requirements, including Eq. (\ref{nullcondition}) and (\ref{nullcounting}) on the defining data $K$, $\{p_i\}$ and $\{q_i\}$ do not depend on the number of layers $L$, and the number of condensed null particles $n_i^{m}$ is always equal to $dim(\mathcal{K})/2$. The 3D states constructed this way is stable in the thermodynamical limit and the remaining degrees of freedom under the condensation will not increase as $L\rightarrow \infty$. Secondly, these layer constructed 3D states do have topologically non-trivial properties. It is straightforward to see (from Eq. (\ref{nullcondition})) that, for the open boundary condition, the particle set $\{q_i\}$ ($\{p_i\}$) is deconfined on the top (bottom) surface as in shown in Fig. \ref{nullparticle} (b). Such 3D states can host 2D topological order on its surface, which is similar to the Walker-Wang model with a modular tensor category\cite{walker2012,simon2013}. Later, we will see that depending on the choice of $\{p_i\}$ and $\{q_i\}$, we can also have deconfined particles in the 3D bulk that organize themselves to form 3D topological orders. In the following sections, we will introduce rigorous results and examples of 3D topological state using the layer construction.

\section{3D states with purely surface topological order}
\label{pSTO}
In this section, we will first introduce a series of examples of 3D states with purely surface topological order (and no bulk topological order). Then we will discuss the general criteria for the construction of 3D states with purely surface topological orders and a Hamiltonian formalism for their surface topological orders.

\subsection{Example: Coupled Layers of the $Z_p$ Toric Code with Non-Trivial Surface Topological Order}
\label{EGpSTO}
\begin{figure}[tb]
\centerline{
\includegraphics[width=3
in]{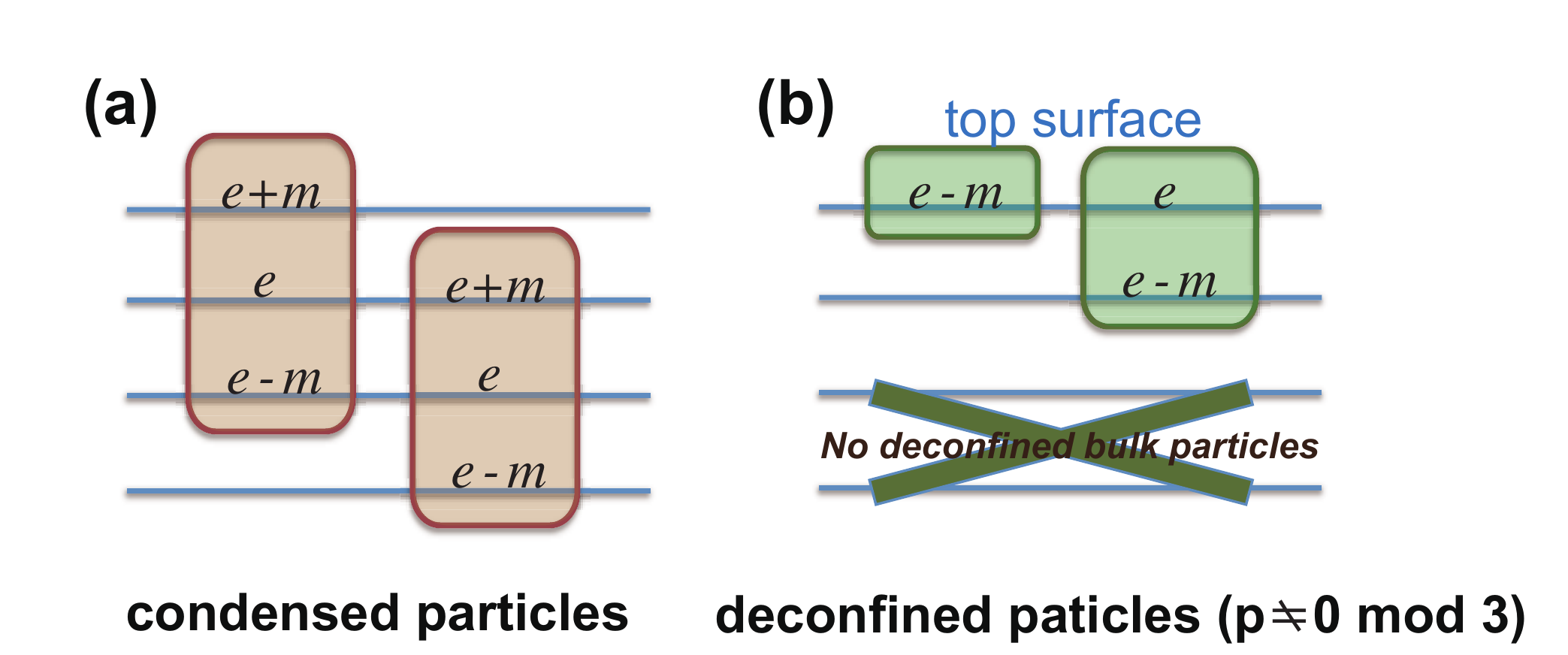}
}
\caption{\label{chiralsurface} (a) Illustration of the condensation of the composite particle of $e+m$, $e$ and $e-m$ in three consecutive layers, with each layer a $Z_p$ toric code state. (b) For $p\equiv 1,2 \text{ mod }3$, there is no deconfined particle in the 3D bulk. The deconfined particles only stay on the surface, which are generated by  
two elementary particles shown in the green circles. 
(The case of $p\equiv 0 \text{ mod }3$ is discussed later in Fig. \ref{coexTopo}.)
}
\end{figure}

Ref. \onlinecite{wang2013} provides a layer constructed model of 3D bosonic topologial insulartor with time-reversal invariant gapped surface states. Motivated by this example, we study the generalization of it and obtain a series of layer constructed models with interesting topologically ordered surfaces. Let's consider the stacked layers of $Z_p$ toric code theory. Each layer of $Z_p$ toric code theory can be described by the K-matrix theory\cite{Kitaev2003, wen2003,ReadSachdev1991} with
\begin{align}
K_{Z_p}=\left(\begin{array}{cc}
0 & p \\ p & 0
\end{array}\right).
\end{align}
There are two types of elementary quasi-particles in the $Z_p$ toric code: the electric particle $(1,0)^T$ (which will also be labelled as the $e$ particles) and the magnetic particle $(0,1)^T$ (which will also be labelled as the $m$ particles). In the stacked layer theory, the null particles we choose to condense take the following form:
\begin{align}
(1,1)^T\otimes e_n + (1,0)^T\otimes e_{n+1} + (1,-1)^T\otimes e_{n+2},
\label{chiralsurfcondensate}
\end{align}
which is the composite particles of $e+m$, $e$ and $e-m$ in three consecutive layers, as is shown in Fig. \ref{chiralsurface} (a). Although we consider composite particles that live in three consecutive layers, we can still recast this type of theory back to the general setting described in Sec. \ref{generalconstrtuction} by viewing two layers of $Z_p$ toric code as one single layer of 2D Abelian topological state. For simplicity of the discussion in this case, we will stick with the current description. One can check that with the particles of the form in Eq. (\ref{chiralsurfcondensate}) condensed, there is no deconfined particle in the 3D bulk when $p\equiv1,2 (\text{mod} 3)$. This can be directly verified by exhausting the possible forms of bulk particles and by noticing that all the particles that are not confined can be viewed as composite particles of the condensed ones. In Sec. \ref{GCpSTO}, we will introduce a more general argument which leads to the same conclusion when applied to this case. When $p\equiv1,2 (\text{mod} 3)$, the only deconfined particles live on the open surface. The surface deconfined particles take the form:
\begin{align}
\alpha_1 (1,-1)^T\otimes e_1 + \alpha_2 ((1,0)^T\otimes e_1+(1,-1)^T\otimes e_2),
\end{align}
where $\alpha_{1,2}=1,2,...,p$. Therefore, we can use a 2-component integer vector $\alpha=(\alpha_1,\alpha_2)$. The two generators of these deconfined surface particles are shown in Fig. \ref{chiralsurface} (b). The topological spin of the particle $\alpha$ can be calculated using the original K-matrix:
\begin{align}
\theta_\alpha &= -\frac{2\pi }{p}(\alpha_1^2+\alpha_1\alpha_2+\alpha_2^2). \nonumber \\
&=\pi \alpha^T M \alpha,
\end{align}
where the second line is written in a matrix form with
\begin{align}
M=-\frac{1}{p}\left(
\begin{array}{cc}
2 & 1 \\ 1 & 2
\end{array}
\right).
\end{align}
The braiding statistics of particle $\alpha$ and $\beta$ is given by
\begin{align}
\theta_{\alpha\beta} &=  -\frac{2\pi }{p}(2\alpha_1 \beta_1+\alpha_1\beta_2+\alpha_2\beta_1+2\alpha_2 \beta_2) \nonumber \\
&=2\pi \alpha^T M \beta.
\label{SurfDeQP}
\end{align}
Since all the surface particles are Abelian, their quantum dimensions are $d_\alpha=1$. There are $p^2$ deconfined particles on the surface. We expect the surface topological order to have total quantum dimension $\mathcal{D}=p$. For a 2D topological state, the chiral central charge $c$, is related to the quantum dimension and topological spins by the following formula \cite{Zhenghan2010}:
\begin{align}
\mathcal{D}e^{\frac{2\pi i c}{8}}=\sum_\alpha d_\alpha e^{i\theta_\alpha}.
\label{centralcharge}
\end{align}
Therefore, we can calculate the chiral central charge for these surface topological orders with $\mathcal{D}=p$:
\begin{align}
c\equiv0 (\text{mod}8),~~~\text{for } p\equiv1 (\text{mod} 3), \\
c\equiv4 (\text{mod}8),~~~\text{for } p\equiv2 (\text{mod} 3).
\end{align}
From this result, we see that, for $p\equiv2 (\text{mod} 3)$, the surface topological state has to have chiral topological order. This result is quite non-trivial because we start with layers of $Z_p$ toric code that are non-chiral. 
Similar situation can also be found using the Walker-Wang construction \cite{walker2012, simon2013}. Also, for the special case with $p=2$, the resulting chiral surface topological state is the so-called $Z_2$ three-fermion state with chiral central charge $c=4$. When realized in $2D$, the corresponding the K-matrix is the Cartan matrix of $SO(8)$:
\begin{align}
K_{SO(8)}=\left(
\begin{array}{cccc}
2 &-1& -1& -1 \\ -1 & 2 & 0& 0 \\ -1 & 0 & 2& 0 \\-1 & 0 & 0& 2
\end{array}
\right).
\end{align}
This specific state is also a realization of the symmetry preserving surface topological state of a time-reversal invariant bosonic topological insulator\cite{wang2013}. To our knowledge, the chiral topological order for the cases with $p>2$ are not well-studied before.

The case with $ p\equiv0 (\text{mod} 3)$ requires a more careful treatment. The set of quasi-particles that are deconfined on the surface is still given by Eq. (\ref{SurfDeQP}), but the naive application of Eq. (\ref{centralcharge}) will fail in this case. That is because other than the trivial particle $\alpha=(0,0)$, there are two other particles $(p/3,p/3)$ and $(2p/3,2p/3)$ that do not have any non-trivial braiding statistics with any other surface deconfined particles. Therefore, the surface theory is not modular if we include all $p^2$ particles label by $(\alpha_1,\alpha_2)$ with $\alpha_{1,2}=1,...,p$. The simplest way to remedy this problem is to take the quotient of the $p^2$ particles by the three ``local" particles $(0,0)$, $(p/3,p/3)$ and $(2p/3,2p/3)$. Then, we end up with a theory of $p^2/3$ particles, namely the total quantum dimension $\mathcal{D}=p/\sqrt{3}$. Using the formula Eq. (\ref{centralcharge}), we obtain that
\begin{align}
c\equiv -2 (\text{mod}8),~~~\text{for } p\equiv0 (\text{mod} 3).
\end{align}
Again, we see that the surface topological order is chiral. For the simplest case with $p=3$, the surface topological order of the quotient theory is the time reversal copy of the $U(1)_3$ Chern-Simons theory with the K-matrix given by
\begin{align}
K^{\text{surf}}_{p=3}=-\left(
\begin{array}{cc}
2 & 1 \\ 1 & 2
\end{array}
\right).
\end{align}
In fact, rather than just a mathematical trick, the quotient does carry physical meanings. The particles that are mod out from the surface theory are a subgroup of particles generated by $(p/3,p/3)$. Written in the stacked-layer theory language, $(p/3,p/3)$ takes the form:
\begin{align}
(2p/3,-p/3)^T\otimes e_1+ (p/3,-p/3)^T\otimes e_2.
\label{neutralparticle}
\end{align}
In fact, we notice that the particles $(2p/3,-p/3)^T\otimes e_n+ (p/3,-p/3)^T\otimes e_{n+1}$ for $\forall n$ are deconfined. Therefore, we should identify them as one  type of bulk deconfined point particle (at different positions in the $z$ direction). This deconfined bulk particle will become the one of the constituents of the 3D bulk topological order. Therefore when we consider the surface topological order, this type of particles should not be included. In the next section, we will come back to this example to study its bulk topological order. To summarize this series of examples, we have obtain the surface topological order with
\begin{align}
&\mathcal{D}=\frac{p}{\sqrt{3}} \text{ and } c\equiv -2 (\text{mod}8),~~~\text{for } p\equiv0 (\text{mod} 3); \nonumber \\
&\mathcal{D}=p \text{ and } c\equiv0 (\text{mod}8),~~~\text{for } p\equiv1 (\text{mod} 3); \nonumber \\
&\mathcal{D}=p \text{ and } c\equiv4 (\text{mod}8),~~~\text{for } p\equiv2 (\text{mod} 3).
\end{align}
For $p\equiv1,2 (\text{mod} 3)$, the 3D bulk is a trivial 3D gapped state. For $p\equiv0 (\text{mod} 3)$, the bulk state has 3D topological order. It would be interesting to work out the effective K-matrix for the surface topological states for all $p$, especially the chiral ones, but this is beyond the scope of this paper and will be left for future works.


\subsection{General Criteria for 3D States with Purely Surface Topological Order}
\label{GCpSTO}
In Sec. \ref{EGpSTO}, we've given a series of examples (for $p\equiv1,2 (\text{mod} 3)$) of 3D states with 2D surface topological order and trivial 3D bulk topological order using the layer construction. In this part of the discussion, we will provide a general criteria for the construction of 3D states with purely surface topological order and trivial bulk topological order.

We denote the quasi-particle lattices generated by $\{p_i\}$, $\{q_i\}$ and $\{p_i\}\cup\{q_i\}$ as $\Gamma_{\{p_i\}}$,$\Gamma_{\{q_i\}}$ and $\Gamma_{\{p_i\}\cup\{q_i\}}$, and the local-particle lattice generated by the column vectors of the K-matrix $K$ as $\Gamma_K$. Notice that the lattice of all quasi-particles is given by the integer lattice $\mathbb{Z}^r$. Then we have the following theorem:

\begin{itemize}
\item  If $\Gamma_{\{p_i\}\cup\{q_i\}}=\mathbb{Z}^r \textnormal{ modulo }\Gamma_K$ and the overlap of $\Gamma_{\{p_i\}}$ and $\Gamma_{\{q_i\}}$ is trivial, i.e. $\Gamma_{\{p_i\}} \bigcap \Gamma_{\{q_i\}}=\emptyset \textnormal{ modulo }\Gamma_K $, the resulting 3D states from the layer construction have only 2D surface topological order and trivial 3D topological order.
\end{itemize}

One can check that the examples discussed in the previous subsection satisfy the condition of this theorem when $p=1,2$ and, therefore, only admits purely surface topological order. The proof of this theorem is the following. Suppose we start with a stack of $L$ layers and turn on the coupling between the layers (assuming open boundary condition in the $z$ direction)  to form a 3D gapped state, as described in Sec. \ref{generalconstrtuction}. As is discussed in Sec. \ref{generalconstrtuction}, the 3D states from the layer construction always havej deconfined surface particle if we consider open boundary condition. Therefore we only need to prove that there is no non-trivial deconfined excitation in the bulk and thus no bulk topological order. If there is a deconfined particle living in the $m_1^\text{th}$ layer and the $m_2^\text{th}$ layer in the bulk, we can always denote it as $X_{m_1,m_2}= \sum_{m=m_1}^{m_2} x_m \otimes e_m$ with $1 \leq m_1 \leq m_2 \leq L$. Without the loss of generality, we can also assume that in this notation $x_m\neq 0$. There are 4 possible situations that need to be discussed separately: i) $1 < m_1 \leq m_2 < L$; ii) $1 = m_1 \leq m_2 < L$; iii) $1 < m_1 \leq m_2 = L$ and iv) $(m_1,m_2)=(1,L)$. We are going to rule out the 4 possibilities one by one.

For the situation i), if there exist a deconfined particle $X_{m_1,m_2}$ with $1<m_1,m_2<L$ for the open boundary (in the $z$ direction) system, this particle will stay deconfined even when one introduces the the coupling between the top and bottom layers by the condensation of the composite particles $p_i\otimes e_L+ q_i\otimes e_1$, $i=1,...,r/2$. However, since $\Gamma_{\{p_i\}} \bigcap \Gamma_{\{q_i\}}=\emptyset \textnormal{ modulo }\Gamma_K $, all condensed null particles $\{n_i^{(m)}\}_{i=1,..,r/2}^{m=1,...,L}$ for the periodic system are linearly independent. The number of these condensed null particles is equal to $dim(\mathcal{K})/2$. Thus, from the result in Ref. \onlinecite{levin2013ac,maissam2013ac1,maissam2013ac2}, there should be no deconfined particle (other than the condensed ones) left in the condensate. Therefore, deconfined particles $X_{m_1,m_2}$ with $1<m_1,m_2<L$ should not exist.

\begin{figure}[tb]
\centerline{
\includegraphics[width=1.5
in]{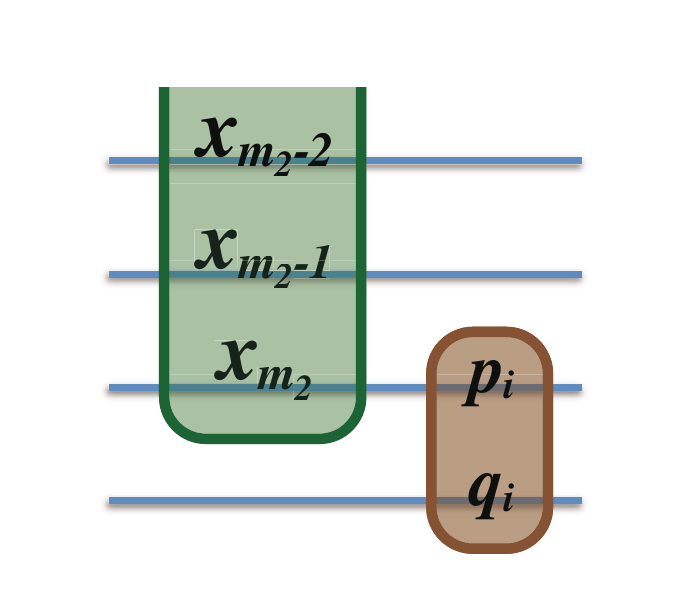}
}
\caption{\label{braidproof} In the discussion of situation ii) (see text), we consider the braiding between a deconfined particle $X_{1,m2}$ (green circle) with a condensed null particle $n_i^{(m_2)}$ (brown circle).
}
\end{figure}

For the situation ii), we consider deconfined particles of the form $X_{1,m2}$ with $m_2<L$. The trivial statistics between $X_{1,m_2}$ and the condensed particles, especially $n_i^{(m_2)}$ (see Fig. \ref{braidproof}), implies that $x_{m_2}$, as a single layer particle, braids trivially with all the $p_i$'s. Since $\Gamma_{\{p_i\}\cup\{q_i\}}=\mathbb{Z}^r \textnormal{ modulo }\Gamma_K$, we can expand $x_{m_2}$ using in the basis of $\{p_i\}\cup\{q_i\}$, namely $x_{m_2}=\sum_i (a^{(m_2)}_i p_i + b^{(m_2)}_i q_i)$.

We will show that the $\sum_i a^{(m_2)}_i p_i$ part of this expansion is 0 modulo local particles. Since $x_{m_2}$ is a single layer particle that has trivial braiding statistics with all the $p_i$'s and all the $q_i$'s also braids trivially with the $p_i$ particles in the single theory (see Eq. (\ref{nullcondition})), the single layer particle $\sum_i a^{(m_2)}_i p_i$ should also braid trivially with all the particle $p_i$'s. Therefore, from Eq. (\ref{nullcondition}), we notice that the particles $(\sum_i a^{(m_2)}_i p_i)\otimes e_m$ with $\forall m$ are deconfined even for the system with periodic boundary condition. As is explained in the previous situation, this is impossible unless $(\sum_i a^{(m_2)}_i p_i)\otimes e_m$'s are local particles.

 Given this, we can consider the fusion between the particle $X_{1,m_2}$ with the condensed particle $-\sum_i b^{(m_2)}_i n_i^{(m_2-1)}$, which does not change the nature of $X_{1,m_2}$ in the condensate. Now notice that $X_{1,m_2}-\sum_i b^{(m_2)}_i n_i^{(m_2-1)}$ is effectively a particle that lives in between the $1^\text{st}$ and $m_2-1^\text{th}$ layers, therefore can be identified as $\tilde{X}_{1,m_2-1}$. By iterating this argument, we can conclude that all deconfined quasi-particles for the situation ii) are equivalent to quasi-particles $X_{1,1}$ living at the top surface combined with a chain of condensed particles. Further, all the deconfined particles $X_{1,1}$ have to be decomposed into linear combinations of $q_i$'s on the top surface. Using similar analysis, we can also show, for the situation iii), that the only possible non-trivial deconfined quasi-particles for the situation ii) are the $X_{L,L}$'s and, therefore, are quasi-particles that live only on the bottom surface. Also, all the deconfined particles $X_{L,L}$ have to be decomposed into linear combinations of $p_i$'s on the bottom surface.

For the situation iv), we focus on the deconfined particles of the form $X_{1,L}$. As is discussed above, we can decompose the $L^\text{th}$ layer component $x_L$ of the $X_{1,L}$ into $x_{L}=\sum_i (a^{(L)}_i p_i + b^{(L)}_i q_i)$, so there exists a deconfined particle $X'_{1,L-1}$ such that $X_{1,L}=X'_{1,L-1}+ \sum_i b^{(L)}_i n_i^{(L-1)} +  \sum_i a^{(L)}_i p_i \otimes e_L$. One can view this as the definition of $X'_{1,L-1}$. The deconfinement of $X'_{1,L-1}$ follows from the fact that  $X_{1,L}$ and $\sum_i a^{(L)}_i p_i \otimes e_L$ deconfines in the open system, and $\sum_i b^{(L)}_i n_i^{(L-1)}$ is condensed. For the deconfined particle $X'_{1,L-1}$, the discussion on the situation ii) already shows that, modulo the condensed particles, $X'_{1,L-1}$ should be identified as $\tilde{X}'_{1,1}$, and therefore as a deconfined particle on the top surface. Thus, the deconfined particle $X_{1,L}$ can always be identified as a trivial composition of the a top surface deconfined particle $\tilde{X}'_{1,1}$ and a bottom surface deconfined particle $\sum_i a^{(L)}_i p_i \otimes e_L$.

Combining the discussion from situation i) to iv), we conclude that, when $\Gamma_{\{p_i\}} \bigcap \Gamma_{\{q_i\}}=\emptyset \textnormal{ modulo }\Gamma_K $ and $\Gamma_{\{p_i\}\cup\{q_i\}}=\mathbb{Z}^r \textnormal{ modulo }\Gamma_K$, the layer constructed 3D state only admits non-trivial deconfined quasi-particles on its surface, and therefore only admits surface topological order. The absence of deconfined bulk quasi-particle indicates the trivialness of the topological properties in the 3D bulk. Also, the discussion of situation ii) and iii) further shows that, with open boundary condition, the only possible excitation on the top (bottom) surface is given by the particle set $\Gamma_{\{q_i\}}$ ($\Gamma_{\{p_i\}}$). Therefore, the surface topological order on the top (bottom) surface is completely determined by the set $\Gamma_{\{q_i\}}$ ($\Gamma_{\{p_i\}}$) and the braiding statistics within it.

\subsection{Surface Topological Orders on Different Surfaces}
\label{surfaceOrder}
Having introduced the general criteria for 3D states with purely surface topological order from the layer construction, we will study the surface topological order on the top, bottom and, especially, the side surface in greater details in the following. As a method to build $3$D gapped states, the layer construction is strongly anisotropic. This anisotropy will manifest itself when we consider a layer-constructed system with finite size in both $z$ and $y$ directions (Remember that the $x$ and $y$ directions are equivalent in our description). The side surface appears to be significantly different from the top and bottom surfaces. We will show that, despite of the superficial distinction among this surfaces, the surface topological orders on them are, in fact, equivalent. The implication of this result will be that the 3D states resulted from the layer construction is topologically equivalent to an isotropic 3D gapped phase. Our discussion will still focus on the case with purely surface topological order.

We start by comparing the surface topological order on the top and bottom surfaces for the system with open boundary along the $z$ direction. As is discussed in Sec. \ref{GCpSTO}, quasi-particles of the surface topological order is given by $\Gamma_{\{q_i\}}$ for the top surface and by $\Gamma_{\{p_i\}}$ for the bottom surface. From the null condition $-p^T_i K^{-1} p_j= q^T_i K^{-1} q_j, \forall i,j$ in Eq. (\ref{nullcondition}), we can establish a one-to-one correspondence between the sets $\Gamma_{\{q_i\}}$ and $\Gamma_{\{p_i\}}$, such that the braiding statistics in the set $\Gamma_{\{q_i\}}$ is exactly the conjugate of that in $\Gamma_{\{p_i\}}$. The braiding statistics discussed here is calculated with respect to a certain normal direction of the layers, which is the $+z$ direction in this case. The $+z$ direction is a natural choice of the normal direction of the top surface, but is opposite to the normal direction of the bottom surface induced by the bulk. Consequently, taking the change of normal direction into account, we need to calculate the braiding statistics using the opposite K-matrix for the two surfaces. Therefore, the null condition
\begin{align}
-p^T_i K^{-1} p_j= q^T_i K^{-1} q_j, \forall i,j
\end{align}
is exactly the condition for the equivalence between the top and bottom surface topological orders.

Having establish the equivalence of top and bottom surfaces, we can move onto the discussion of the surface topological order on the side surface. First, let's briefly review the edge theory of a 2D Abelian topological state with a K-matrix $K$ and its behavior under quasi-particle condensation. The edge state of a 2D Abelian topological state with the K-matrix $K$ is described by the $1+1$D chiral Luttinger liquid theory\cite{wen1990edge}:
\begin{align}
\mathcal{L}_{\text{CL}}=\frac{K_{IJ}}{4\pi} \partial_x \phi^I \partial_t \phi^J- V_{IJ} \partial_x \phi^I \partial_x \phi^J,
\end{align}
where $V$ the velocity matrix. Suppose the 2D bulk is in the condensed phase of a set of particles $\{n_i\}$ that satisfies $n_i^T K^{-1} n_j=0, \forall i,j$. The condensation is induced by turning on extra coupling terms in chiral Luttinger liquid on the edge\cite{levin2013ac,maissam2013ac1,maissam2013ac2}:
\begin{align}
\delta\mathcal{L}_\text{cond}= \sum_i -g_i \cos(c_i n_i^T \phi),
\end{align}
where $g_i$'s are the coupling constants and $c_i\in \mathbb{Z}$ is the minimal integer, for each $i$, such that $e^{i c_i n_i^T \phi}$ only create/annihilate a local particle. Here we've already organized the $\phi_I$'s into a column vector $\phi$. Since $e^{i c_i n_i^T \phi}$ is a local operator, the $\delta\mathcal{L}_\text{cond}$ only involves local terms and, therefore, can be written as a local coupling using the microscopic degrees of freedom. Notice that the commutation relations between $n_i^T\phi$'s (with different $i$'s) are trivial in the chiral Luttinger liquid theory:
\begin{align}
[n_i^T\phi, n_j^T\phi]=0,~~~\forall~i,j.
\end{align}
Therefore, in the deep condensation limit with $g_i\rightarrow \infty$, the coupling term $\delta \mathcal{L}_\text{cond}$ force the fields $n_i^T\phi$ to develop non-zero vacuum expectation values:
\begin{align}
\langle n_i^T\phi\rangle \neq 0.
\end{align}
If the number of fields $n_i^T\phi$ equals $\text{dim}(K)/2$, then edge theory becomes fully gapped in the condensed phase.

Now we will use this language to describe the topological order on the side surface of the layer-constructed 3D state with purely surface topological order. Before we introduce the condensation that couples the layers, the side surface of the stacked layer is described by the Lagrangian density:
\begin{align}
\mathcal{L}_{\text{CL}}^\text{side}=
\sum_m\frac{1}{4\pi} \left(\partial_x \Phi^{(m)}\right)^T K \left( \partial_t \Phi^{(m)} \right) \nonumber \\
-\left(\partial_x \Phi^{(m)}\right)^T V \left( \partial_x \Phi^{(m)} \right),
\label{SurfChiralLuttinger}
\end{align}
where $\Phi^{(m)}$ is the $r$-component (remember $r=\text{dim}(K)$) chiral boson field on the edge of the $m^\text{th}$ layer and $V$ is the velocity matrix for each layer. By quantizing this theory, we obtain the commutation relation of the boson fields:
\begin{align}
[\partial_x \Phi^{(m)}_I(x), \Phi^{(n)}_J(y)]=-i2\pi (K^{-1})_{IJ} \delta(x-y) \delta_{n,m}.
\end{align}
The condensation of the particles $\{n^{(m)}_i\}$ that couple the layers also introduces the coupling in the chiral edge mode:
\begin{align}
\delta\mathcal{L}_\text{cond}= \sum_{i,m} - g_{m,i} \cos\left(l_i (p_i^T \Phi^{(m)} +q_i^T \Phi^{(m+1)})\right),
\label{SurfCoupling}
\end{align}
where $g_{m,i}$'s are the coupling constants, and $l_i\in \mathbb{Z}$ the minimal integer which ensure the locality of the operator $e^{il_i (p_i^T \Phi^{(m)} +q_i^T \Phi^{(m+1)})}$. That is to say $e^{il_i (p_i^T \Phi^{(m)} +q_i^T \Phi^{(m+1)})}$ is the creation/annihilation operator of a local particle. This means that the layer construction can, in principal, be implemented by coupling the microscopic degrees of freedom without using non-local terms.

From the canonical quantization of the chiral Luttinger theory Eq. (\ref{SurfChiralLuttinger}) and the null condition Eq. (\ref{nullcondition}), we see that the fields in the $\cos$ terms have trivial commutation relations:
\begin{align}
[(p_i^T \Phi^{(m)} +q_i^T \Phi^{(m+1)}),(p_j^T \Phi^{(n)} +q_j^T \Phi^{(n+1)})]=0,~~\forall i,j,m,n.
\end{align}
Therefore, at the strong coupling limit $g_{m,i}\rightarrow \infty$, the fields $(p_i^T \Phi^{(m)} +q_i^T \Phi^{(m+1)})$ will develop non-trivial vacuum expectation value, namely
\begin{align}
\langle p_i^T \Phi^{(m)} +q_i^T \Phi^{(m+1)} \rangle \neq 0.
\end{align}
In the following, we will denote the field $p_i^T \Phi^{(m)} +q_i^T \Phi^{(m+1)}$ as $\Lambda_i^{m+1/2}$.

\begin{figure}[t]
\centerline{
\includegraphics[width=3.5
in]{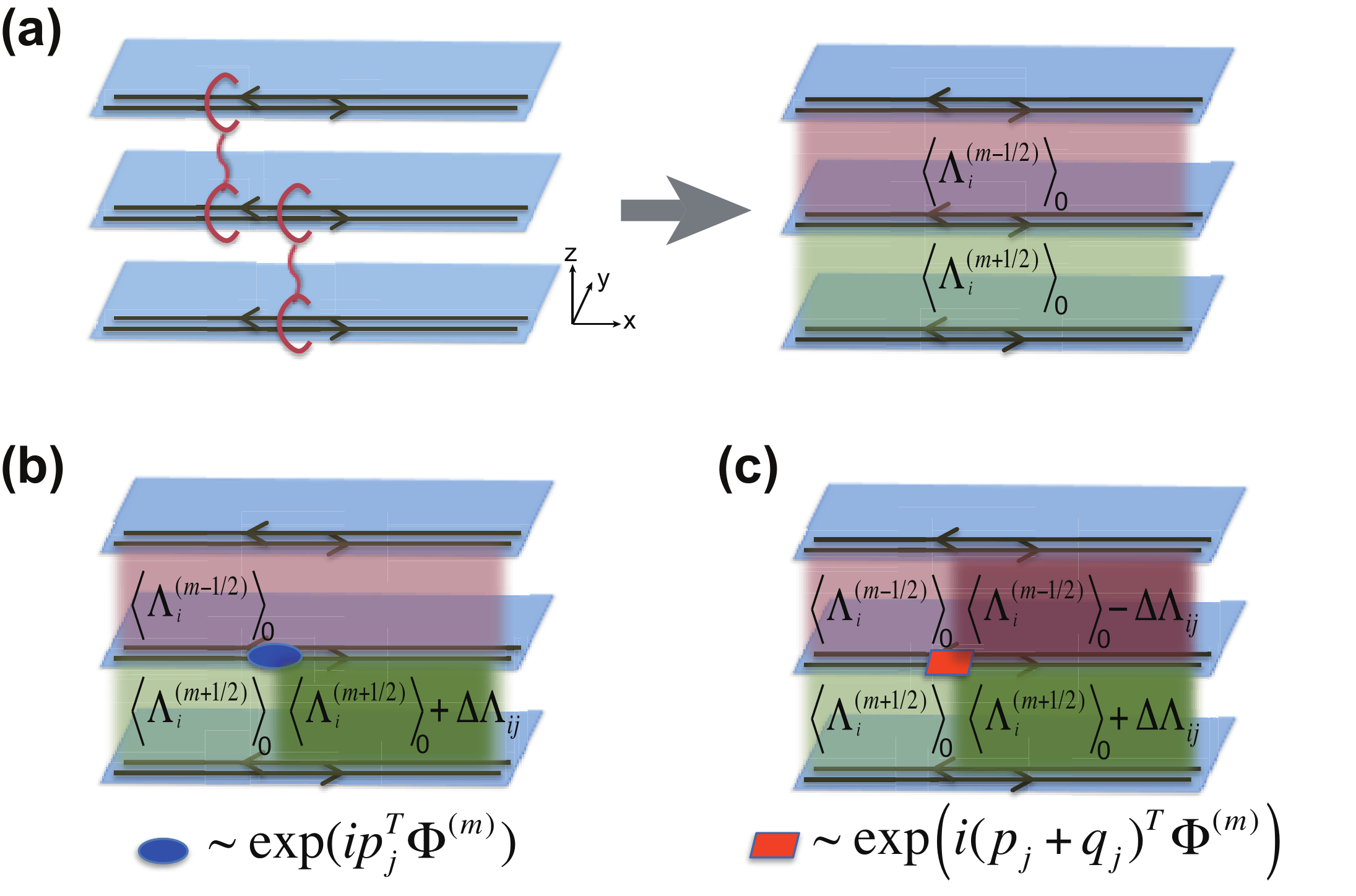}
}
\caption{\label{sideSurf} (a) The condensation in the 3D bulk induces a coupling between the edge states of each layer. The side surface is effectively a coupled wire system. In the strong coupling limit, the vacuum expectation values $\langle \Lambda_i^{m+1/2}\rangle_0$ are pinned to the minima determined by the $\cos$ terms in Eq. (\ref{SurfCoupling}). The coloring on the right hand side indicates non-trivial, but uniform $\langle \Lambda_i^{m\pm 1/2}\rangle_0$ on the side surface. (b) The operator $e^{i p^T_j{\Phi^{m}(x)}}$ creates a collection of kinks (depicted as the change in color) in $\langle \Lambda_i^{m+1/2}(x)\rangle$, which will be identified as a topological quasi-particle $w_j$ on the $m+1/2^\text{th}$ layer of the side surface. (c) The operator $\chi_j^m$ creates a kink-anti-kink pair
that effectively tunnels
the quasi-particle $w_j$ from the $m-1/2\text{th}$ layer to the $m+1/2\text{th}$ layer.
}
\end{figure}

Now we can use this formalism to study the topological order on the side surface. From the Lagrangian density in Eq. (\ref{SurfChiralLuttinger}) and Eq. (\ref{SurfCoupling}), we notice that we are in fact solving a problem of coupled array of 1D wires.  Similar coupled wire systems that exhibits 2D topological order were studied in Ref. \onlinecite{sondhi2001, kane2002, teo2014}. The building blocks of these systems are 1D wires of normal Luttinger liquids. For our problem, we are considering the coupled arrays of fractionalized 1D wires which are the edge states of the 2D Abelian topological orders, as is shown in Fig. \ref{sideSurf} (a).

\begin{figure}[t]
\centerline{
\includegraphics[width=2.5
in]{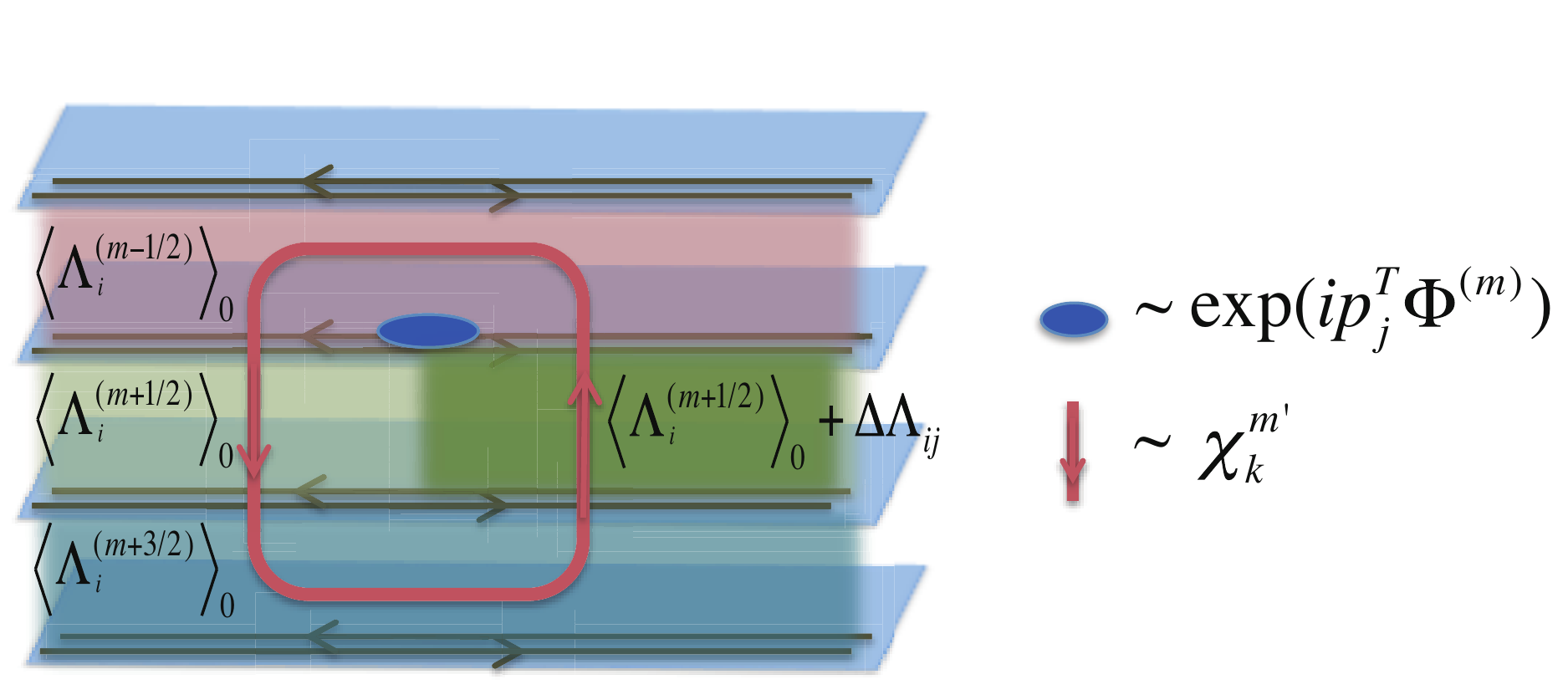}
}
\caption{\label{surfbraid} This figure illustrates the braiding between two quasi-particles of the surface topological order on the side surface. The blue dot represent the quasi-particle $w_j$ created by the operator $e^{ip_j^T \Phi^(m)}$. The red line represents the trajectory of the second quasi-particle $w_k$ that the first one braids with. The operators that tunnel $w_k$ along the trajectory is explained in the main text. Especially, the vertical tunnel operator from the ${m'}^\text{th}$  layer to the ${m'+1}^\text{th}$ is given by $\chi_k^{m'}$.
}
\end{figure}

To identify the topological order on the side surface, we need to identify the non-trivial quasi-particles and their statistics. Deep in the condensed phase, the fields $\Lambda_i^{m+1/2}$ only fluctuate very weakly around its vacuum expectation value which are pinned by the $\cos$ terms in Eq. (\ref{SurfCoupling}). In fact, the $\cos$ terms create multiple degenerate minima for the fields $\Lambda_i^{m+1/2}$. The domain walls, or kinks, of $\langle \Lambda_i^{m+1/2}\rangle$ between different minima are topologically stable excitations and will be identified as topological quasi-particles on the side surface. We can start from the configuration with all $\langle \Lambda_i^{m+1/2}\rangle_0$ uniform. The operator $e^{i p^T_j{\Phi^{(m)}(x)}}$ creates a series of kinks in $\langle \Lambda_i^{m+1/2}\rangle$, $i=1,...,r/2$ (see Fig. \ref{sideSurf} (b)), such that the vacuum expectation values of $\langle \Lambda_i^{m+1/2} (x)\rangle$ on the right and left sides of $e^{i p^T_j{\Phi^{(m)}(x)}}$ differ by
\begin{align}
\Delta\Lambda_{ij}=2\pi p_i^T K^{-1} p_j.
\end{align}
We will identify the $e^{i p^T_j{\Phi^{(m)}(x)}}$ as the creation operator of a quasi-particle, denoted $w_j$, on the side surface. Since the kinks created by $e^{i p^T_j{\Phi^{m}(x)}}$ are the domain walls of $\langle \Lambda_i^{m+1/2} (x)\rangle$, we would like to think of this quasi-particle $w_j$ as residing in the $m+1/2^\text{th}$ layer. The quasi-particle $w_j$ can tunnel between the $m-1/2^\text{th}$ layer and $m+1/2^\text{th}$ layer, which is implemented by the tunneling operator:
\begin{align}
\chi_j^{m}=e^{i (p^T_j+q^T_j){\Phi^{m}(x)}}.
\end{align}
The operator $\chi_j^{m}$ creates simultaneously a kink in $\langle \Lambda_i^{m+1/2} (x)\rangle$ and an anti-kink in $\langle \Lambda_i^{m-1/2} (x)\rangle$ as is shown in Fig. \ref{sideSurf} (c). It is straightforward to show that, in the strong coupling limit,
\begin{align}
\chi_j^{m+1}  e^{i p^T_j{\Phi^{(m)}(x)}} = e^{i \langle \Lambda_j^{m+1/2} (x)\rangle} e^{i p^T_j{\Phi^{(m+1)}(x)}},
\label{TunnelPhase}
\end{align}
which means that the operator $\chi_j^{m}$ tunnels the quasi-particle $w_j$ from the $m-1/2^\text{th}$ layer to the $m+1/2^\text{th}$ layer and, more importantly, with a phase factor $e^{i \langle \Lambda_j^{m+1/2}(x)\rangle} $. We will show that this phase factor will give rise to the braiding statistics of the quasi-particles on the side surface. We can consider the configuration with a quasi-particle $w_j$ on the $m+1/2^\text{th}$ layer created by the operator $e^{ip_j^T\Phi^{(m)}(x)}$, as is shown in Fig. \ref{surfbraid}. A close path of the quasi-particle $w_k$ is depicted as the red line. The horizontal tunneling of the quasi-particle $w_k$ in the uniform backgrounds of $\langle \Lambda_i^{m-1/2} (x)\rangle_0$ and $\langle \Lambda_i^{m+3/2} (x)\rangle_0$ does not produce any non-trivial phase factor. The vertical tunneling between the layers through the operator $\chi^{m'}_k$'s, as is shown in Eq. (\ref{TunnelPhase}), does carry non-trivial phase factors. Due the existence of the kinks in $\langle \Lambda_i^{m+1/2} (x)\rangle$, the net phase, which is viewed as the braiding phase between the quasi-particle $w_j$ and $w_k$, is non-zero and is given by:
\begin{align}
B_{jk}= e^{-i\Delta\Lambda_{jk}} =e^{-2\pi i p_j^T K^{-1} p_k}.
\end{align}
Therefore, by identifying the quasi-particle $w_i$ on the side surface with the deconfined particle $p_i$ on the bottom surface, we conclude that topological order on the side surface is the same as the one on the bottom surface. Furthermore, by noticing that the operators $e^{ip_j^T\Phi^{(m)}(x)}$ and $e^{-iq_j^T\Phi^{(m+1)}(x)}$ create exactly the same kink configuration on the side surface, we can also show the equivalence of the side surface and the top surface through the same analysis.

\section{3D Topologically Ordered States}
\label{3DTO}
In this section, we will focus on the layer construction of 3D topologically ordered states. In Sec. \ref{GCpSTO}, we have introduced the general criteria for the construction of a 3D gapped state with trivial bulk topological order. When this criterion is not satisfied, we will generically have bulk deconfined excitations, which will organize themselves to form  3D topological order. Unlike the topological order in 2D which is fully characterized by the modular tensor category theory, the general structure of 3D topological order is not yet known completely. The layer construction provides a method to construct 3D topological ordered gapped states that will shed light on the potential structure for a generic 3D topological order. In this section, we will consider three types of different examples of three dimensional topological order. Firstly, we will introduce the construction of conventional topological order with particle-string mutual statistics that resembles the $Z_p$ lattice gauge theory in 3 spatial dimensions. Secondly, we will construct an example with 3D topological order with coexisting surface topological order. Thirdly, we will consider a more non-trivial example with braiding statistics not only between particles and strings but also between different types of strings.

\subsection{Conventional 3D Topological Order}
\label{Convent3DTO}
\begin{figure}[t]
\centerline{
\includegraphics[width=3.5
in]{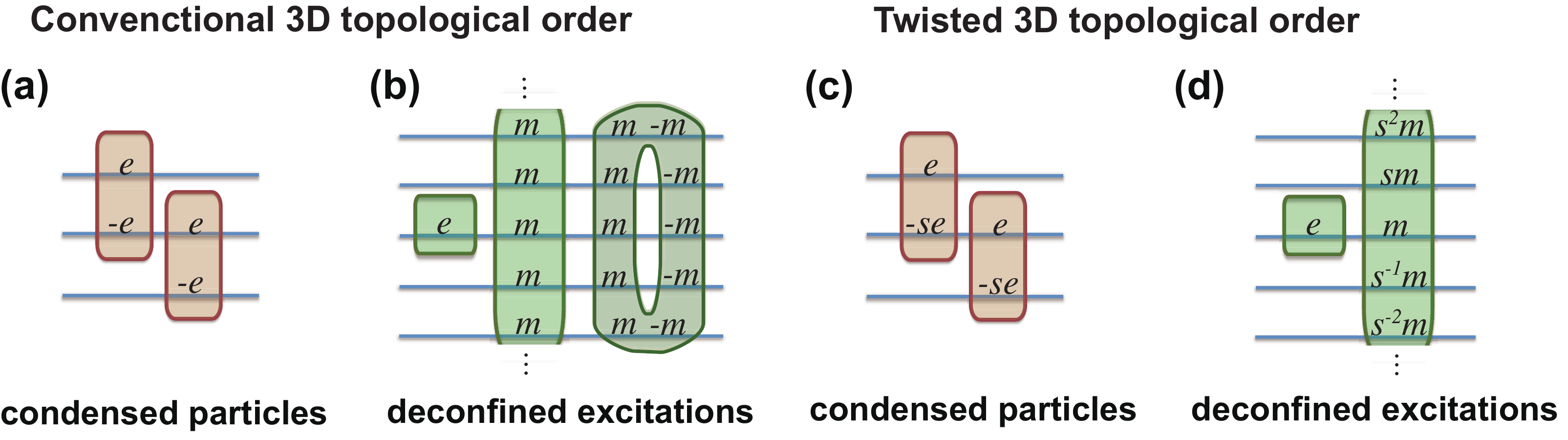}
}
\caption{\label{LGT3d} (a) The condensation of $e$ and $-e$ composite in the two consecutive layers in the system of stacked layers of $Z_p$ toric codes. (b) The electric particle in each layer is a deconfined particle that should be identified as the electric particle in the 3D $Z_p$ lattice gauge theory. The string of $m$ particles in every layer is a deconfined string excitation which can be viewed as the flux string in the $Z_p$ gauge theory. The flux strings don't have to align strictly along the $z$-direction. Contractible flux string of finite size, as is shown in the figure, can be considered. (c) A different set of condensed particles in layers of $Z_p$ toric code, the condensation of which leads to the twisted lattice gauge theory (see text). (d) The deconfined particles in the twisted theory, including an electric particle and the ``twisted" flux string. 
}
\end{figure}

To construct 3D gapped state with the conventional 3D topological order of 3D $Z_p$ lattice gauge theory, we start with layers of $Z_p$ toric code with the K-matrix:
\begin{align}
K_{Z_p}=\left(
\begin{array}{cc}
0 & p \\ p & 0
\end{array}
\right).
\end{align}
We choose
\begin{align}
p_1=(1,0)^T,~~~q_1=-(1,0)^T.
\end{align}
The condensed particles are illustrated in the Fig. \ref{LGT3d} (a). This condensation can be understood as an ``exciton condensate", in which the electric particle in each layer, i.e. $(1,0)\otimes e_m ~\forall m$, deconfined.
\begin{align}
\epsilon_e = (1,0)^T\otimes e_m.
\end{align}
This electric particle $\epsilon_e$ can hop between the layer through the fusion with the condensed ``excitons". Since spin of the electric particle $\epsilon_e $ is $0$, it is a boson in the 3D system. Besides the deconfined electric particle, the other type of deconfined excitation is the string excitation which is composed of the magnetic particle $m$ in the each layer, i.e
\begin{align}
\epsilon_{m\text{-string}}=\sum_m (0,1)^T\otimes e_m,
\end{align}
as is shown in Fig. \ref{LGT3d} (b). So far, we've only considered the string excitations that orient along the $z$-direction. String excitations can in fact take, especially in the continuum limit, any orientations with their excitation energy proportional to their length. We can consider contractible strings of finite size as depicted in Fig. \ref{LGT3d} (b).
From the K-matrix, it is easy to see that the braiding statistics between the electric particle and the magnetic string is
\begin{align}
\theta_{e,m-\text{string}}=2\pi/p.
\label{EMstat}
\end{align}
This statistical phase is independent from the location of the point-like particle $\epsilon_e$. By comparing with the 3D $Z_p$ lattice gauge theory, the deconfined electric particle can be identified as the electric particle in the $Z_p$ gauge theory, while the magnetic string can be identified as the magnetic flux string. These two types of deconfined excitation and their mutual statistics give rise to the topological order in 3D. In this example, the layer construction studies the anisotropic limit of the $Z_p$ gauge theory in which the $m$-string always orient itself along the $z$-direction. 
The string excitation can actually take arbitrary shape, and its energy is always linearly proportional to its length. The statistics between the point particles and string excitations should not depend on the shape of the string. Therefore, the layer construction faithfully captures the topological properties of 3D lattice gauge theory.

Now, we can consider a variation of this case with a topological twist in each layer. For the convenience of the discussion, we will assume that $p$ is a prime number such that $Z_p$ is a field, namely $s^{-1}\in Z_p, \forall s \in Z_p $. In the following, we will use $s^{-1}$ to denote the integer whose product with $s$ is 1 modulo $p$. For a given $s\in Z_p$, we consider the condensate defined by
\begin{align}
p_1=(1,0)^T,~~~q_1=-(s,0)^T.
\end{align}
The condensed particles in this case are shown in Fig. \ref{LGT3d} (c). It is easy to see that the $e$ particle in each layer is still deconfined. It will be identified as a type of point particle in the 3D state. The hopping of $e$ particles between layers is still implemented through the fusion with the condensed particles. We see that in this hopping process an $e$ particle of the $m^\text{th}$ layer turns into a $se$ particle of $m+1^\text{th}$ layer. We can understand this process as a hopping of the $e$ particle followed by a symmetric twist. To be more precise, for $Z_p$ toric code theory, the topological properties of the theory, namely the braiding statistics, fusion rules, are invariant under the map generated by $e\rightarrow se, ~m\rightarrow s^{-1}m$. This mapping is a symmetry of the $Z_p$ toric code as far as the topological properties are concerned. We will call this symmetry action as the symmetric twist. The hopping process in this case is the hopping of the $e$ particle under one layer to the other followed by the symmetric twist $e\rightarrow se, ~m\rightarrow s^{-1}m$.

Based on this understanding, one should expect that the topological order obtained in the case is the same as the untwisted case. Indeed, in this twisted case, there is a twisted version of the $m$-string (see Fig. \ref{LGT3d} (d)) which is given by
\begin{align}
\epsilon'_{m\text{-string}}=\sum_m (0,s^{-m})\otimes e_m.
\end{align}
We pick the unit of point-like excitation and string-like excitation to be the ones shown in Fig. \ref{LGT3d} (d). Their mutual braiding statistics is given by
\begin{align}
\theta'_{e,m-\text{string}}=2\pi/p.
\end{align}
This is the same as the untwisted case. Also, we see that the hopping of the $e$-particle between the layers is compatible with its braiding with the twisted $m$-string. So far, the analysis indicates that the local properties of the twisted case are the same as those of the untwisted one. However, these analysis is done for an infinite system or a system with open boundaries. When we consider periodic boundary condition in the $z$-direction, the twisted case automatically generates a monodromy when we consider the hopping of the electric particle along the non-trivial cycle along the $z$-direction. To be more precise, an electric particle $e$ becomes the particle $s^L e$ after traversing along the $z$-direction cycle, where $L$ is the number of layers in the system. The twist generated by the condensed particle particle can be undone locally by considering the symmetric twist generated by $e\rightarrow s^{-l}e$ and $m\rightarrow s^{l}m$ for the $l^\text{th}$ layer. Then the system with a periodic boundary condition is equivalently described by the stacked layers of $Z_p$ toric code with following condensed particles: $(1,0)^T\otimes e_m+(1,0)^T\otimes e_m+1$ with $m=1,..,L-1$ and $(s^{-L},0)^T\otimes e_L+(1,0)^T\otimes e_1$. The system is locally equivalently to the untwisted case, and the twist along the $z$-direction cycle is generated by the coupling between the $1^\text{st}$ and the
$L^\text{th}$ layer induced by the condensation of the particle $(s^{-L},0)^T\otimes e_L+(-1,0)^T\otimes e_1$. Therefore we can think of the twisted model as the conventional 3D $Z_p$ gauge theory with 2D membrane twist defect between the $1^\text{st}$ and the $L^\text{th}$ layer. (The notation of twist defects in 2D and their properties are first studied in Ref. \onlinecite{KongTD2012,maissamTD2013}. Its generalization in 3D is first studied in Ref. \onlinecite{YingTD2013}.) This membrane defect implements the symmetry twist $e \rightarrow s^L e$ in the 3D $Z_p$ gauge theory. Let's look at the ground state degeneracy (GSD) of this twisted model on $T^3$. Assume that we also have periodic boundary conditions along $x$ and $y$ directions. For the untwisted case, we can study the Wilson loop operators $W_{x}$, $W_{y}$ and $W_{z}$ that tunnel the $e$ particle around the non-trivial cycles along the $x$, $y$ and $z$ directions. The Wilson loop operators measure the magnetic flux in the non-trivial cycles and can take $p$ different values $e^{i2\pi n/p}$, $n=1,...,p$ for $Z_p$ gauge field. Therefore, the untwisted system has $p^3$ degenerate ground states on $T^3$. For the twisted case, $W_z$ is not well-defined because the twist membrane render the tunneling a cycle around the $z$ direction un-closed unless $s^L\equiv 1 (\text{mod }p)$. $W_x$ and $W_y$ are still well-defined. However, if we adiabatically move the tunneling path of $W_{x(y)}$ all together one cycle along the $z$ direction, the twist membrane generates a map: $W_{x(y)} \rightarrow W^{s^L}_{x(y)} $. This implies that the flux along the $x(y)$ direction measured by $W_{x(y)}$ should be invariant under map, and therefore can only take the flux value 1 unless $s^L\equiv 1 (\text{mod }p)$. Therefore, when $s^L\not\equiv 1 (\text{mod }p)$, the GSD on $T^3$ is 1. For $s^L\equiv 1 (\text{mod }p)$, the twist membrane only perfoms a trivial twist which means the GSD on $T^3$ is the same as the untwisted case. These results are summaried as follows:
\begin{align}
&\text{GSD on } T^3=1, \text{ for } s^L\not\equiv 1 (\text{mod }p), \nonumber \\
&\text{GSD on } T^3=p^3, \text{ for } s^L\equiv 1 (\text{mod }p).
\end{align}
Therefore, we see that the twisted model has a different behavior of the ground state degeneracy on $T^3$ from the untwisted case.

\subsection{Coexisting Bulk and Surface Toploogical Order}
\label{Coexist}
\begin{figure}[t]
\centerline{
\includegraphics[width=2.5
in]{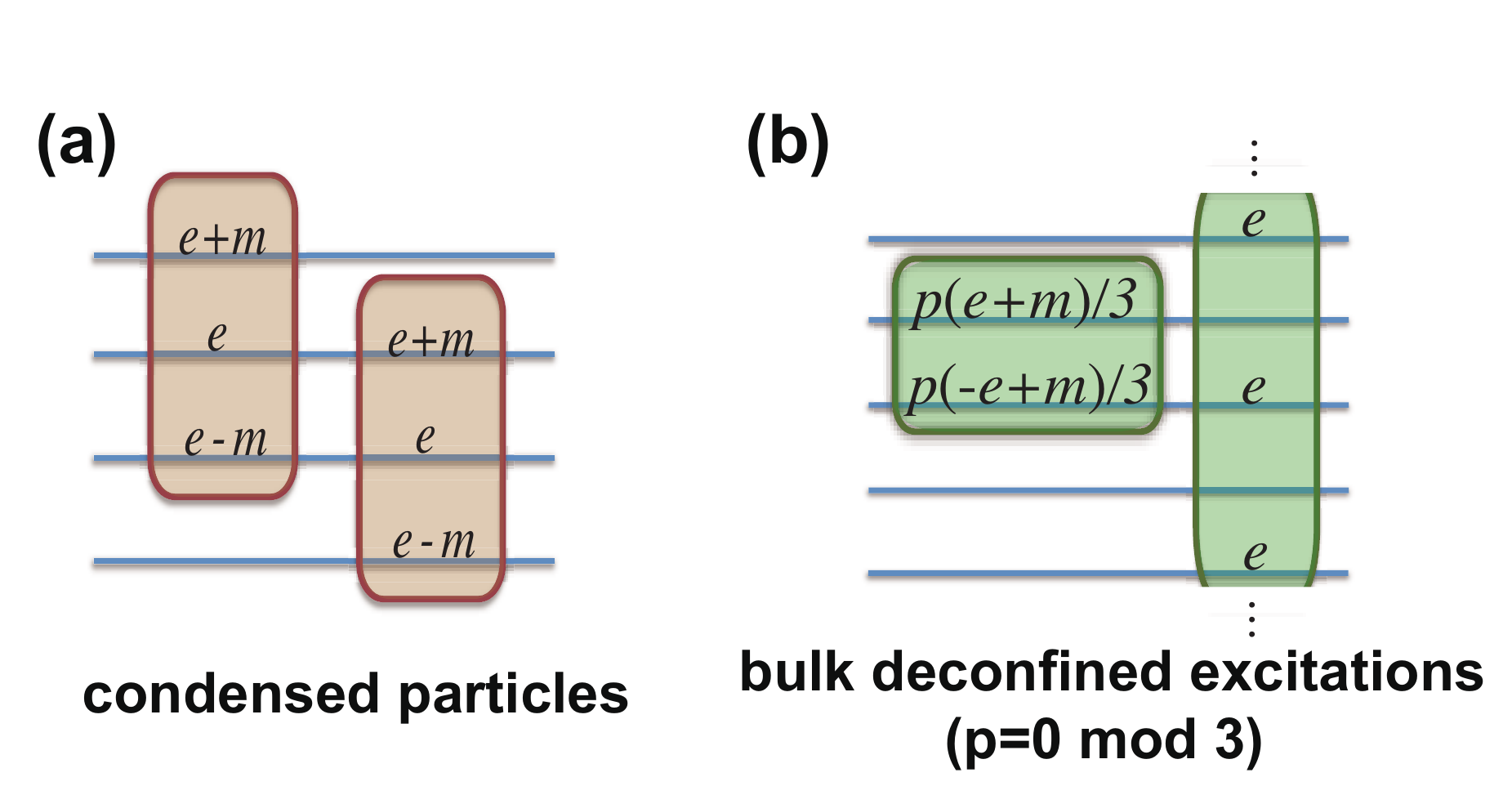}
}
\caption{\label{coexTopo} (a) The condensed particle in the coupled $Z_p$ toric code system. This case is the same as the case studied in Sec. \ref{EGpSTO}, but here we focus on the cases with  
$p\equiv 0 (\text{mod}3)$. (b) The bulk deconfined excitations in this sytem. The one on the left is a point particle while the one on the right is a string excitation. They have non-trivial mutual statistics.
}
\end{figure}

In Sec. \ref{EGpSTO}, we discussed a construction using layers of $Z_p$ toric codes with the condensed particles given by Eq. (\ref{chiralsurfcondensate}) (see Fig. \ref{chiralsurface} (a)). As is explained there, for $p$ dividable by $3$, 
there are two particles which have trivial statistics (except the vacuum particle) with all the rest of surface deconfined particles. One of them can be written as
\begin{align}
(p/3,p/3)^T \otimes e_1 + (-p/3,p/3)^T \otimes e_2.
\end{align}
The other particle is a two-particle bound state of this one. (This form is essentially equivalent to the expression given in Eq. (\ref{neutralparticle}).) After the quotient of all the surface deconfined particles by these two particles, we can consistently calculate the central charge of the surface topological order. This implies that these two trivial particles shouldn't be considered as part of the surface topological order. In fact, one can check that particles of the form:
\begin{align}
\epsilon_\text{pt}=(p/3,p/3)^T \otimes e_m + (-p/3,p/3)^T \otimes e_{m+1}
\end{align}
are all deconfined. Therefore, this particle should be viewed as a bulk deconfined point particle, which is a generalization of the bulk deconfined electric particle in the previous example. Again, the spin of the particle $\epsilon_\text{pt}$ is 0, so it is a boson in the 3D system. This is consistent with the fact that any point particle in 3D should have either bosonic or fermionic statistics. Similar to the previous example, there is also a type of deconfined  string excitation that takes the simple form of
\begin{align}
\epsilon_\text{str}=\sum_m (1,0)^T \otimes e_{2m},
\end{align}
which is shown in Fig. \ref{coexTopo} (b). Naively, there is another type of string-like deconfined excitation that take the form of $\sum_m (1,0)^T \otimes e_{2m+1}$. However, it can be identified as $\epsilon_\text{str}$ fused with a string of condensed particles and, therefore, is considered as topologically equivalent to $\epsilon_\text{str}$. The braiding statistical angle between the deconfined point particle and the string excitation is
\begin{align}
\theta_\text{pt,str}=2\pi/3.
\end{align}
Therefore, the bulk topological order of this case is equivalent to that of a 3D $Z_3$ lattice gauge theory. However, as is discussed in Sec. \ref{pSTO}, this system also hosts a surface topological order with chiral central charge $c\equiv 2 (\text{mod} 3)$ on its open boundary. Therefore, it is a layer-constructed system that has co-existing bulk and surface topological order.

This situation can also be compared with the Walker-Wang model\cite{walker2012}. The input data of the Walker-Wang model is a braided tensor category. If the braided tensor category (which describes a 2D topological order) is modular, namely only the vacuum particle in this tensor category braids trivially with all the rest of the quasi-particles, the Walker-Wang model provides a construction of a 3D gapped state with trivial bulk topological order but non-trivial surface topological order characterized by the input modular tensor category.
If the input data is only pre-modular, there will be, by definition, transparent particles (other than the vacuum particle) which braids trivially with all other particles. In the Walker-Wang model associated to a pre-modular tensor category, the transparent particles can propagate in the bulk and become the quasi-particles of the 3D topologically ordered states. This scenario seems similar to the case of our layer constructed model. For this model with $p\equiv1,2(\text{mod}3)$, the surface topological order studied in Sec. \ref{pSTO} is modular and the 3D bulk has trivial topological order. For $p\equiv0(\text{mod}3)$, the deconfined particles on the surface only forms a pre-modular tensor category. If we quotient out the surface particles that braid trivially with others, we get a consistent surface topological order. Those particles indeed penetrate in the bulk and form 3D topological order. Due to the similarity between these two situations, one might want to conjecture that the lattice model realizations of the layer construction in these cases can be formulated using the Walker-Wang model.

\subsection{More Generic 3D Tolopological States with String Braiding Statistics}
\label{GenericState}
\begin{figure}[t]
\centerline{
\includegraphics[width=3
in]{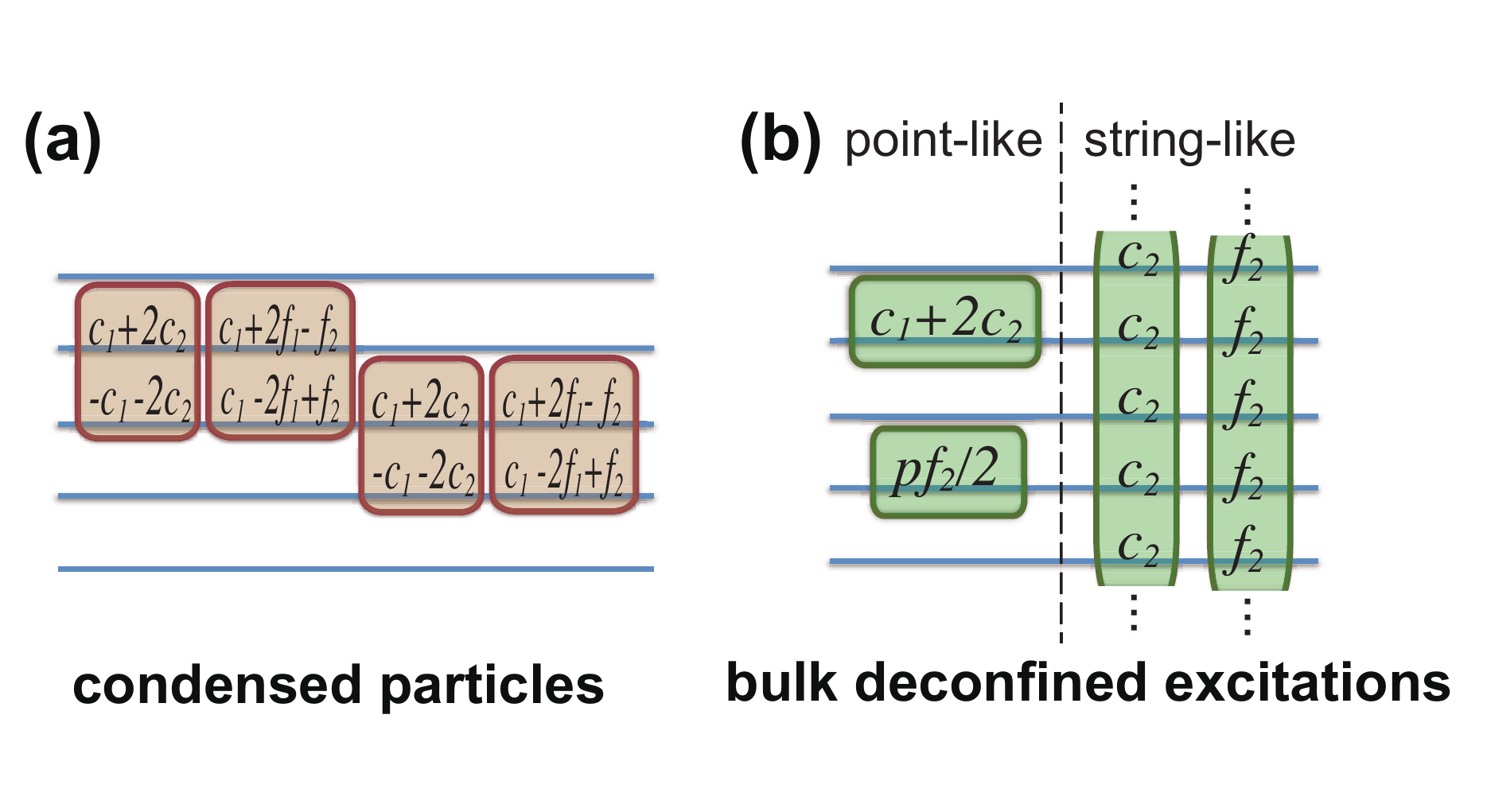}
}
\caption{\label{EGgeneric} (a) The condensed particles for a more generic layer constructed 3D state. Each layer contains two identical copies of $Z_p$ toric code, with $p$ a multiple of $4$. $c_{1,2}$ and $f_{1,2}$ are the charge and flux particles in the two $Z_p$ toric code system, respectively. (b) The set of deconfined excitations, including two types of point particles and two types of string excitations with nontrivial mutual braiding.
}
\end{figure}

Now we will consider a more generic situation. The starting point is still stacked layers of $Z_p$ toric code. Here we will first assume that $p$ is a multiple of 4. For the convenience of later discussions, we view two layers of $Z_p$ toric codes as one single layer described by the K-matrix:
\begin{align}
K_0=\left(\begin{array}{cccc}
0 & p & 0 & 0\\
p & 0 & 0 & 0\\
0 & 0 & 0 & p\\
0 & 0 & p & 0\\
\end{array}
\right)
\end{align}
For later convenience, we will denote the four generators of all quasi-particles $(1,0,0,0)^T$, $(0,1,0,0)^T$, $(0,0,1,0)^T$ and $(0,0,0,1)^T$ as $c_1$, $f_1$, $c_2$ and $f_2$. The condensed particles (see Fig. \ref{EGgeneric}) are given by the data:
\begin{align}
p_1=(1,0,2,0)^T,~~~q_1=(-1,0,-2,0); \nonumber\\
p_2=(1,2,0,-1)^T,~~~q_2=(1,-2,0,1).
\end{align}
In this condensate, there are two types of deconfined point particles
\begin{align}
&\epsilon_\text{c-pt}=(1,0,2,0)^T\otimes e_m, \nonumber \\
&\epsilon_\text{f-pt}=(0,0,0,p/2)^T\otimes e_m,~~~\forall m
\end{align}
as is shown in Fig. \ref{EGgeneric} (b). These point particles can hop between the layers by fusion with condensed particles. Their mutual statistics is trivial. There are also two different types of string excitations (Fig. \ref{EGgeneric} (b)) :
\begin{align}
\epsilon_\text{c-str}=\sum_m (0,0,1,0)^T \otimes e_m, \nonumber \\
\epsilon_\text{f-str}=\sum_m (0,0,0,1)^T \otimes e_m.
\end{align}
The braiding statistics between the point particles and string excitations are given by
\begin{align}
\theta_{I,J}=2\pi\left(
\begin{array}{cc}
2/p & 0 \\ 0 & 1/2
\end{array}
\right),
\label{pt_f-str}
\end{align}
where $I=1,2$ stand for point particles $\epsilon_\text{c-pt}$ and $\epsilon_\text{f-pt}$ respectively while $J=1,2$ stand for string excitations $\epsilon_\text{f-str}$ and $\epsilon_\text{c-str}$.
More interestingly, from the expression of $\epsilon_\text{c-str}$ and $\epsilon_\text{f-str}$, it seems that these string excitations can also have non-trivial mutual braiding statistics. Indeed, if the system is periodic along the $z$-direction, the two types of strings can wrap around the $z$-direction and their mutual braiding phase is given by $\omega_\text{c-str,f-str}=2\pi L/p$, where $L$ is the number of layers in the system. This result can simply be obtained by looking the system from the $z$-direction and viewing it as a quasi-2D system. From this point of view, the two string excitations are simply point particles in the 2D sense. Their braiding is also understood as the braiding of particles in 2D. This analysis only applies to vertical strings that wrap around the $z$ direction cycle. We need to inspect further the string braiding statistics for strings with finite size.

\begin{figure*}[t]
\centerline{
\includegraphics[width=7
in]{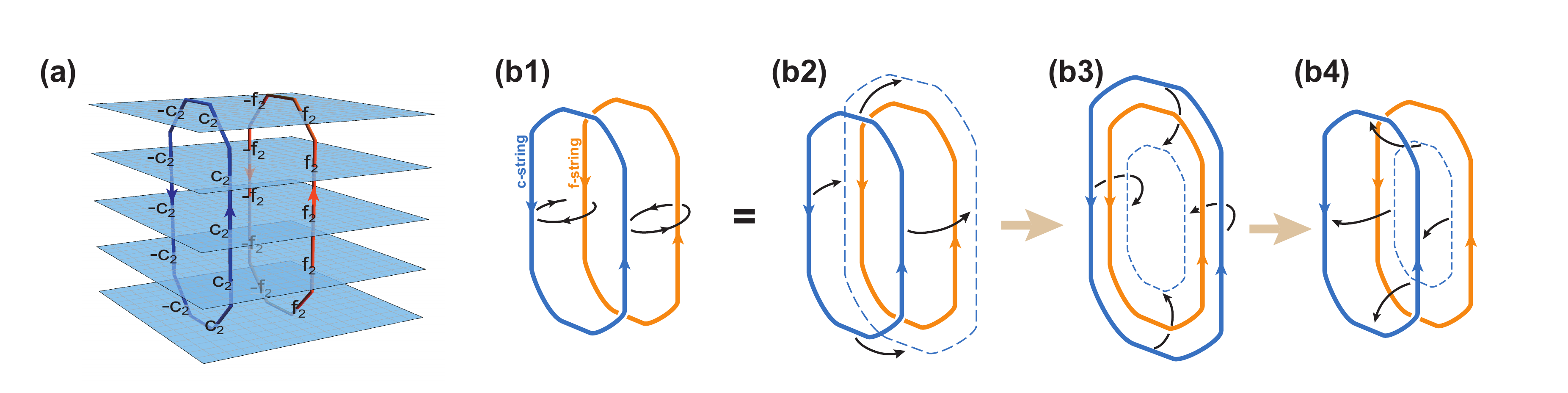}
}
\caption{\label{TwoLoops}  (a) The definition of $c$-string (blue) and $f$-string (orange). Each plane represents a layer of 2D topological state. 
The convention on the orientation of the string is explained in the main text. (b1) The braiding process of the two strings. (b2-b4) The braiding process shown in decomposed steps.
}
\end{figure*}

The strings with finite size can be constructed in the same way as we did in Sec. \ref{Convent3DTO}. Fig. \ref{TwoLoops} (a) illustrates the construction of contractible strings of finite size for the example system. Each plane represents a layer in the layer construction. The particle contents of the string excitations in each layer are shown. For example, the blue string in Fig. \ref{TwoLoops} (a) represents the finite size version of the deconfined excitation $\epsilon_\text{f-str}$, which is given by a line with $c_2$ in each of the $5$ layers together with another line of $-c_2$ in each of the $5$ layers. In general, an oriented string is defined as a chain of particles by the following rule. When the string crosses a plane and the direction of the string is parallel (anti-parallel) to the normal direction of the plane (which is $\hat{z}$ direction in our example), a $c_2$ ($-c_2$) particle is assigned to the crossing point. 
 The same definition applies to the other type of string except for a substitution of $c_2$ with $f_2$. In the following, we will refer to the string consisting of $c_2$ particles as the $c$-string, and the other type as the $f$-string. It is straightforward to see that these string are deformable. The deformation of string along the $xy$ plane is simply the change of position of $c_2$'s or $f_2$'s within the layers. The deformation along $z$-direction is achieved by pair creation or pair annihilation of $c_2$'s or $f_2$'s. For example, if we fuse the $c_2$ and $-c_2$ of $f$-string in the bottom layer in Fig. \ref{TwoLoops} (a), the string effectively shrinks to a smaller size.

\begin{figure*}[t]
\centerline{
\includegraphics[width=7
in]{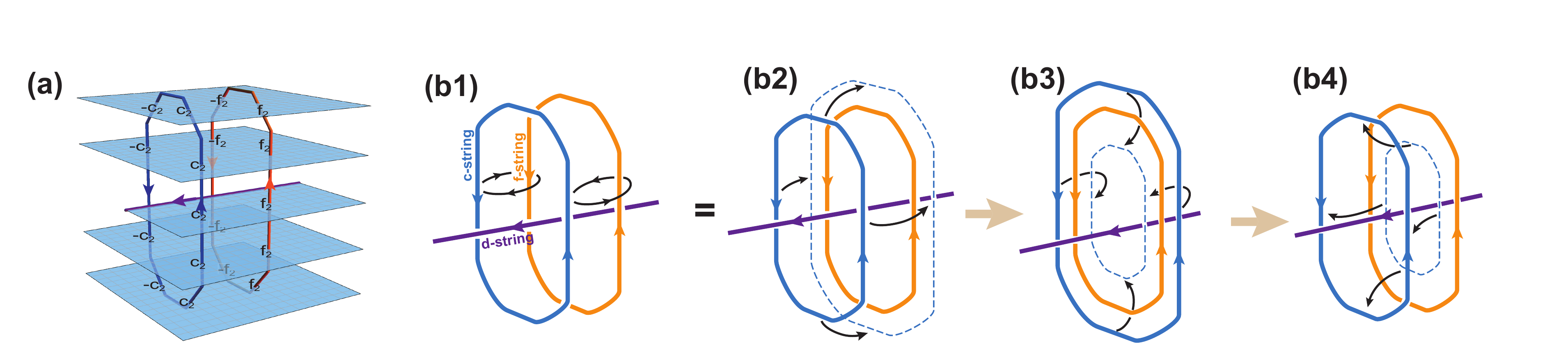}
}
\caption{\label{TwoLoopsDis} (a) The configuration of the $c$-string and the $f$-string that link with an edge dislocation (or a $d$-string) which is indicated by the purple line. (b1) The braiding process of the $c$-string and $f$-string that are both linked with the $d$-string. (b2-b4) show the decomposed steps in this braiding process.
}
\end{figure*}

The braiding between two types of strings is defined to be the process  shown in Fig. \ref{TwoLoops} (b1-b4). Fig. \ref{TwoLoops} (b1) is the overall action to the $c$-string (blue) to complete the braiding with the $f$-string (orange). For the sake of clarity, we've shown the decomposed steps in this action in Fig. \ref{TwoLoops} (b2-b4). First, we expand the $c$-string and move it past the $f$-string such that the $f$-string effectively passes through the ``inside" of $c$-string (Fig. \ref{TwoLoops} (b2)). Then, we shrink the $c$-string and move it back through the inside of the $f$-string (Fig. \ref{TwoLoops} (b3)). Once the $c$-string passes through the $f$-string, we expand the $c$-string back to its original size and move it back to the original position (Fig. \ref{TwoLoops} (b4)). In fact, it is straightforward to show that these type of braiding is always trivial for contractible $c$-string and $f$-string. One can verify this by considering the total phase that results from the braiding of the particles $c_2$'s and $f_2$'s in each layer during the braiding of strings. The braiding phase between all the $c_2$'s with all the $f_2$'s is exactly canceled by that between all the $-c_2$'s and all the $-f_2$'s due to the opposite directions of the two braid. This is consistent with the observation of Ref. \onlinecite{Levin2014StrBrd} that the braiding between two contractible strings must be trivial. 

In order to obtain some non-trivial string braiding statistics, we need to consider a system with an edge dislocation and the $c$-string and $f$-string that are linked with the edge dislocation (See Fig. \ref{TwoLoopsDis} (a)). In the following, we will refer to the edge dislocation with a unit Burgers vector as the $d$-string. The orientation of the $d$-string is determined through the right-hand rule by the normal direction of the surface that ends at the $d$-string. Now, we can study the braiding process between the $c$-string and the $f$-string, both of which are linked to a $d$-string. 
Fig. \ref{TwoLoopsDis} (b1) shows the overall action to the $c$-string (blue) in the braiding process at presence of the $d$-string. Fig. \ref{TwoLoopsDis} (b2-b4) show the decomposed steps for the braid. (Similar three-string braiding processes in $Z_N^k$ or $Z_{N_1}\times Z_{N_2}\times Z_{N_3}\times Z_{N_4}$ gauge theory are considered in Ref. \onlinecite{Levin2014StrBrd, Ran2014StrBrd, JuvenWenStrBrd2014}.) In this process, the string braiding statistics, which is given by the total braiding phase between the quasi-particles that form these string, is
\begin{align}
\omega_{c,f}^d = \theta_{c_2,f_2}= 2\pi/p,
\label{cfbraid}
\end{align}
where $\theta_{c_2,f_2}$ is the braid statistics of the $c_2$ and $f_2$ particles in a singe layer of the 2D topological state defined by the K-matrix $K_0$. More generally, if we consider a composite string of $n_c$ $c$-strings and that of $n_f$ $f$-strings that are both linked with an edge dislocation with Burgers vector of length $b$, the string braiding statistics is given by
\begin{align}
b~n_c n_f  \omega_{c,f}^d = 2\pi n_c n_f b/p.
\label{cfbraid}
\end{align}
From Fig. \ref{TwoLoopsDis} (a), it is easy to see that the net contribution to this phase $\omega_{c,f}^d $ is essentially given by the braiding between the $c_2$ particle and the $f_2$ particle in the defected layer. Compare to the braiding statistics between point particles and string excitations in Eq. (\ref{pt_f-str}), when $p$ is a multiple of $4$, the string braiding statistics $\omega_{c,f}^d $ is more fractionalized. Thus, it cannot be removed by attaching point particles to the strings. Also, the fact that the $d$-string is an extrinsic defect rather than a dynamical one does not undermine the topological protection of the string braiding statistics $\omega_{c,f}^d $ either. That is because throughout the braiding process, the $d$-string does not have to move which means that no additional dynamical phase because of the extrinsic nature of the $d$-string should come in to play. Furthermore, the three-string braiding phase $\omega_{c,f}^d$ can be related to the braiding phase $\omega_\text{c-str,f-str}=2\pi L/p$ between two vertical strings winding around the torus, which we discussed at the beginning of this subsection. 
An $L$ layer system can be formally considered as a $0$ layer system (vacuum) with an edge dislocation in the $xy$ plane at infinity with Burgers vector $L\hat{z}$, {\it i.e.} $L$ $d$-strings at infinity.
With periodic boundary condition, the $c$-strings and the $f$-string that wrap around the nontrivial cycle in the $z$ direction in fact have non-trivial linking number with all of the L $d$-strings, and therefore, have non-trivial string braiding statistics $\omega_\text{c-str,f-str}$ that is proportional to the number of layers $L$.

\begin{figure}[t]
\centerline{
\includegraphics[width=3
in]{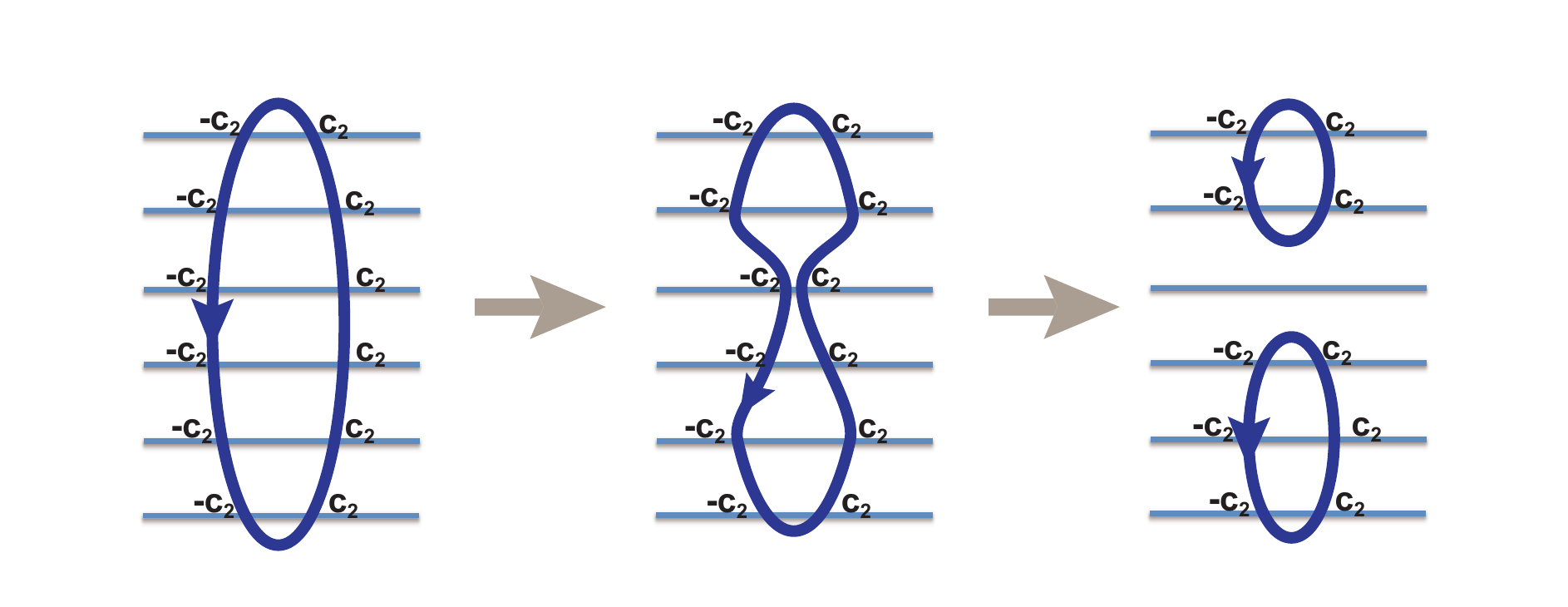}
}
\caption{\label{PinchedLoop} One $c$-string can be deformed and split into two $c$-strings through the fusion of $c_2$'s within the layers. The reverse process can be viewed as the fusion between two $c$-strings.
}
\end{figure}

There is an alternative approach to compute the string-string braiding statistics, which will be helpful for our later discussion. To explain this approach, we consider the string fusion and splitting process in similar way as Ref. \onlinecite{Levin2014StrBrd}. 
When we consider our system as a multi-layer 2D system, the finite-size $c$-string and $f$-string are special configurations of quasi-particles that live in some number of layers. Therefore, their deformation and fusion can be understood in term of the fusion of quasi-particles in the 2D theory. The specific type of deformation that we are interested in is the deformation that change the number of strings. 
Here, we will show that the string excitations in the layer constructed system can also deform, especially split and fuse, without accumulating non-trivial Berry's phase. To demonstrate this, we can consider the example shown in Fig. \ref{PinchedLoop}. One $c$-string can be split into two independent $c$-strings by bringing two different points on the string very close to each other and pinching this point off. In this process, we use nothing more than the fusion rules of quasi-particles in each layer. In the current setting, $c_2$ in each layer is an Abelian particle with no spin. Therefore, the splitting process does not have any non-trivial Berry's phase associated with it. One can also consider the reverse process of splitting, which is the fusion of two $c$-strings. The same analysis is also applicable to the $f$-strings.

\begin{figure*}[t]
\centerline{
\includegraphics[width=7
in]{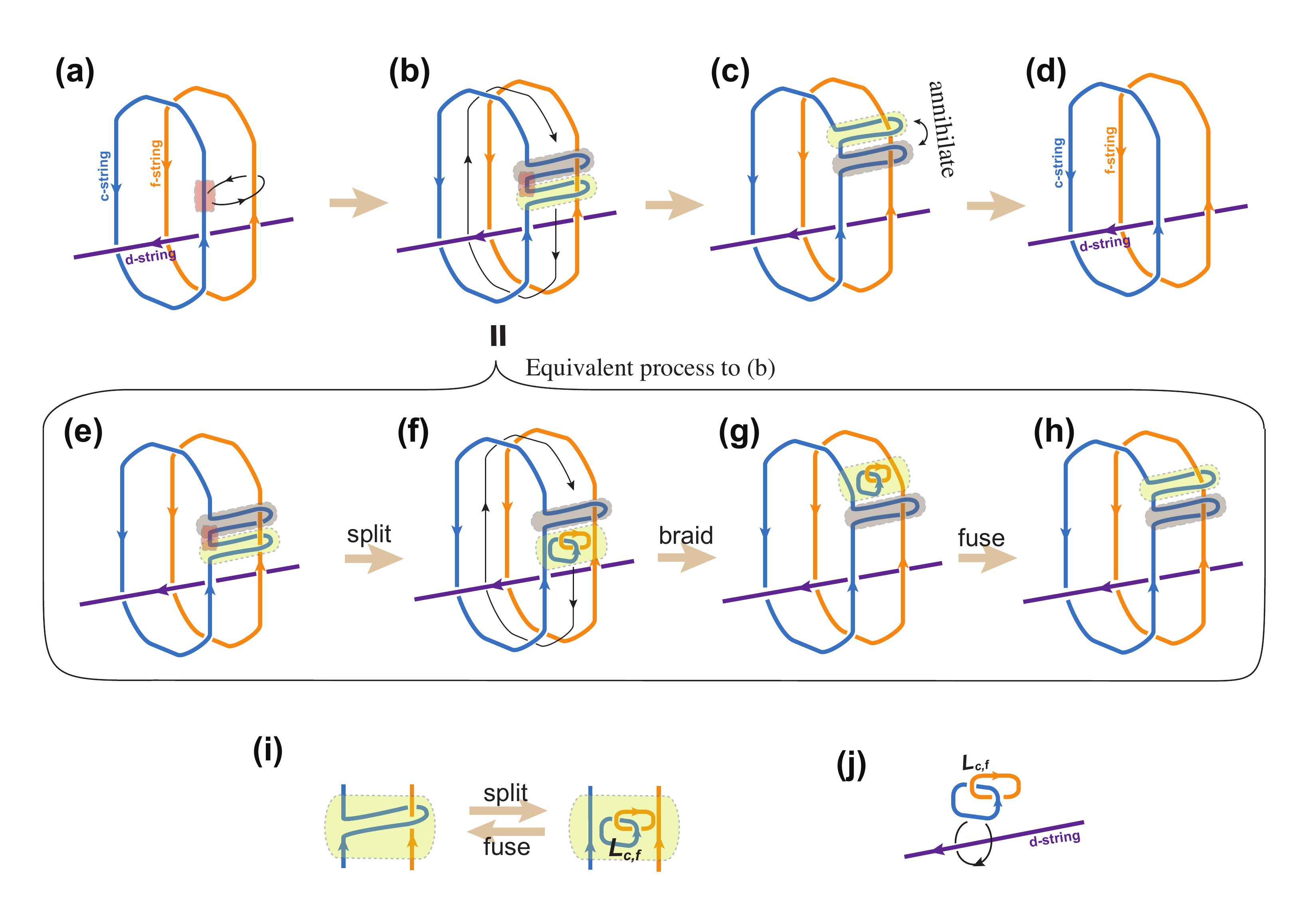}
}
\caption{\label{BraidDeform} (a-d) An alternative string braiding process that is topologically equivalent to that shown in Fig. \ref{TwoLoopsDis}. 
Upon fusion and splitting the strings the process in (b) can be deformed to the combined process shown in (e-h). (i) shows the local identification of string configuration under splitting and fusion which plays a central role in the re-interpretation of the process shown in (b). The total Berry's phase in the string braiding process equals the Berry's phase in the process induced by the braiding of the link $L_{c,f}$ around the $d$-string, as is shown in (j)
}
\end{figure*}

\begin{figure}[t]
\centerline{
\includegraphics[width=3.5
in]{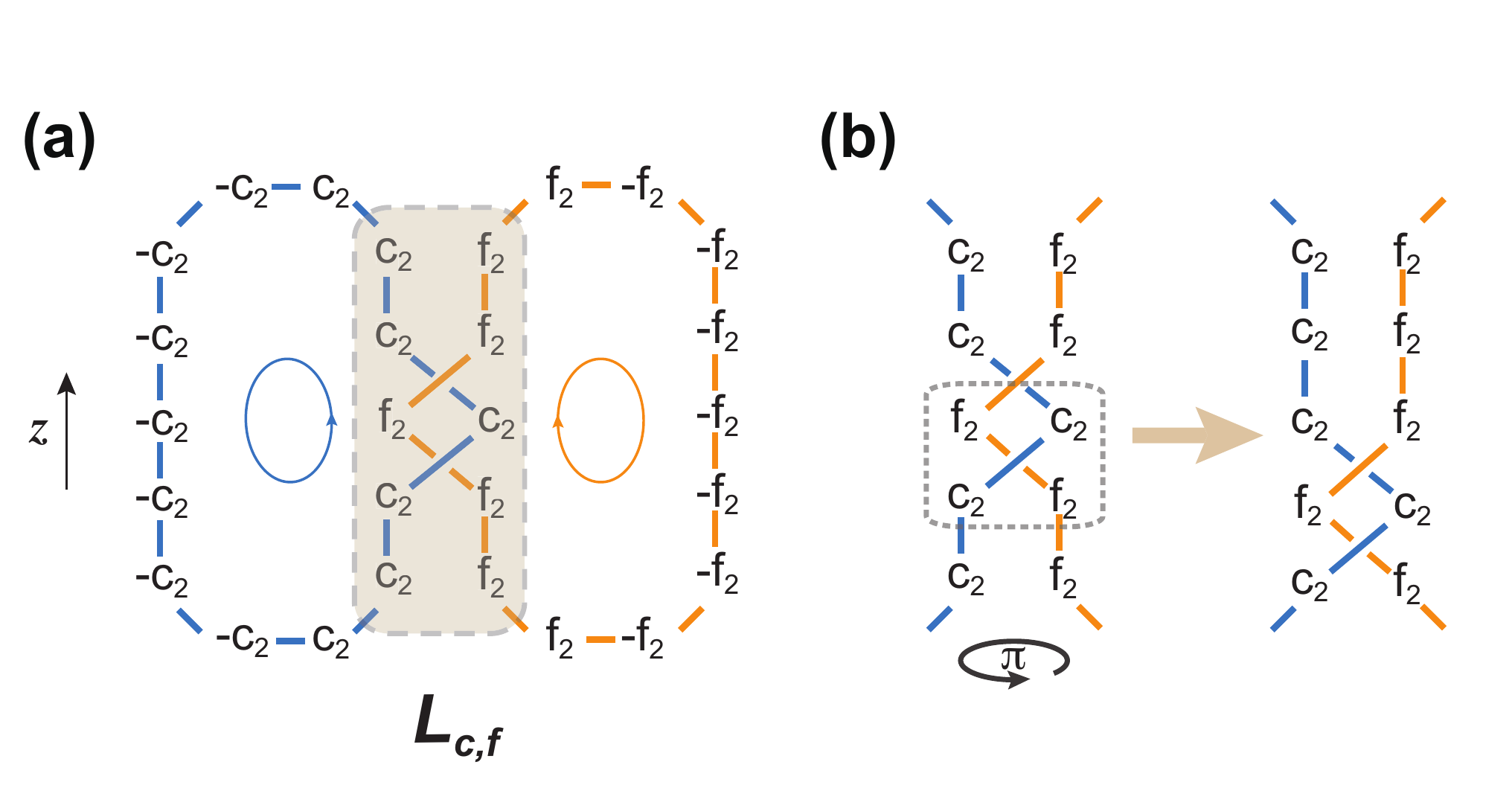}
}
\caption{\label{LinkedMove} (a) The configuration of the link $L_{c,f}$ with the quasi-particles associated with it in each layer. The layers are horizontal which are not drawn in this figure. In this link $L_{c,f}$, the $c$-string and the $f$-string have linking number $1$. Their orientation is indicated by the colored arrow (blue for the $c$-string and orange for the $f$-string). (b) The operation that brings the linked part of the two strings down by one layer along the $-z$ direction is a $\pi$ rotation of the part of the link indicated by the grey dashed box.
}
\end{figure}

Since for each layer of 2D topological state, the fusion and braiding of quasi-particles are operations that commute with each other, we argue that the string braiding also commutes with string deformation including splitting and fusion of strings. To be more precise, we can deform, split and fuse strings without changing the total Berry's phase in the braiding of string as long as the overall process remains topologically equivalent. Using this argument, we can obtain a new view to the braiding process (Fig. \ref{TwoLoopsDis} (b1)) that we discussed above. The braiding process is equivalent to the process shown in Fig. \ref{BraidDeform} (a-d). First, we pick an infinitesimal segment (the red shaded area in Fig. \ref{BraidDeform}(a)) of the $c$-string, and braid it around the $f$-string without moving the rest of the $c$-string. This action creates locally a link (yellow region) and an anti-link (grey region). The link and anti-link can annihilate each other if we bring them close together. Now, instead of annihilation them together directly, we bring the local link in the yellow region around the $d$-string (Fig. \ref{BraidDeform}(b)). We then annihilate the yellow and the grey regions by bringing them together afterwards (Fig. \ref{BraidDeform}(c)). The whole process shown in Fig. \ref{BraidDeform} (a-d) is topologically equivalent to the braiding of strings defined in Fig. \ref{TwoLoopsDis} (b1). We argue that in the combined process of Fig. \ref{BraidDeform} (a-d), only the process in Fig. \ref{BraidDeform} (b) contributes to the non-trivial Berry's phase . This is because 1) The process shown in Fig. \ref{BraidDeform}(a) only moves an arbitrarily small segment and should not accumulate any non-trivial Berry's phase; and 2) The annihilation between the yellow and grey regions also does not contribute to the Berry's phase since it can also be thought of as bringing an infinitesimal segment of the $c$-string around the $f$-string as we did in Fig. \ref{BraidDeform}(a). Given that all the non-trivial Berry's phase comes from the process Fig. \ref{BraidDeform} (b), we will re-interpret it using the fusion and splitting of strings. As is shown in Fig. \ref{BraidDeform} (i), the strings in the yellow region can be deformed, upon splitting and fusion, to unlinked $c$-string and $f$-string accompanied by a link $L_{c,f}$ of linking number 1 between the $c$-string and $f$-string. Therefore, the process shown in Fig. \ref{BraidDeform}(b) is equivalent to the combined process shown in Fig. \ref{BraidDeform}(e-h). Since the fusion and splitting do not produce non-trivial Berry's phase in this process, the total Berry's phase equals to the Berry's phase $\Omega^d_{L_{c,f}}$ of braiding the link $L_{c,f}$ around the the $d$-string. The analysis above can be summarized by the equation:
\begin{align}
\omega^d_{c,f}=\Omega^d_{L_{c,f}}.
\label{BraidEquiv}
\end{align}

In the analysis above, for the sake of generality, we've implicitly assumed that the $c$-strings and $f$-strings can be considered as continuous object in the space. We expect that its validity still holds even when the space is discrete. In fact, we can verify Eq. (\ref{BraidEquiv}) by explicitly calculating $\Omega^d_{L_{c,f}}$ in our layer constructed system. We start by studying the Berry's phase induced by the local motion of the link $L_{c,f}$. If we hold the relative position of the $c$-string and $f$-string in the link $L_{c,f}$ fixed, it is obvious that moving $L_{c,f}$ along the $x$ or the $y$ direction does not produce any non-trivial phase. In contrast, moving the $L_{c,f}$ in the $z$-direction while keeping the relative positions of the $c$-string and the $f$-string fixed will induce non-trivial Berry's phase. The link configuration of $L_{c,f}$ is drawn in Fig. \ref{LinkedMove} (a). The orientation of the strings are indicated by the arrows of the corresponding colored circles (blue and orange). The layers that are supposed to be represented by horizontal lines are omitted in this figure, and the $z$ direction is indicated by the black arrow. Let's consider moving this configuration by one layer down towards the $-z$ direction. For the parts outside grey shaded region in Fig. \ref{LinkedMove} (a), since there is no non-trivial linking between these segments of the $c$-string and $f$-string, they can be deformed into the configuration that is one layer below their original position without inducing any Berry's phase. In the grey shaded region, the linked parts of $c$-string and $f$-string need to be treated more carefully. To move the linked part one layer towards the $-z$ direction, we essentially have to rotation the part of the link enclosed by the grey dashed line in Fig. \ref{LinkedMove} (b) by $\pi$ about the $z$ axis. The net Berry's phase associated to this rotation is equivalent to a full braid between $c_2$ and $f_2$ in a single layer and is, therefore, $\theta_{c_2,f_2}=2\pi/p$. By similar analysis, the Berry's phase in moving the link $L_{c,f}$ one layer up toward the $z$ direction is the opposite. In process of moving the $L_{c,f}$ around the the $d$-string (Fig. \ref{BraidDeform}(g)), namely an edge dislocation, the net effect is to move the $L_{c,f}$ by the Burgers vector, {\it i.e.}, to move it downwards by one layer. Therefore, we can conclude that
\begin{align}
\Omega^d_{L_{c,f}}=\theta_{c_2,f_2},
\end{align}
which verifies the consistency of Eq. (\ref{BraidEquiv}). In this discussion, we always have to keep the $d$-string fixed because it is an extrinsic defect rather than a dynamical excitation. If we can promote the $d$-strings to dynamical excitations, we can study other types of string braiding between the three types of strings $c$, $f$ and $d$, such as $\omega^c_{f,d}$ and $\omega^f_{d,c}$\cite{Levin2014StrBrd}. The model that we considered so far is not capable of capturing the dynamics of the $d$-strings, but the analysis above relating string-string braiding and link-string braiding can be used to derive general identities on the string braiding, as will be discussed in Sec. \ref{GeneralA}. 

\section{Topological Field Theory Description}
\label{FieldTheory}

In this section, we introduce a topological field theory description of the layer-constructed system, which provides a phenomenological description of the particle-string braiding and string-string braiding. The field theory can be considered as a continuum limit of the discrete system of coupled layers, although a generic derivation starting from coupled layers remain an open question that we will leave for future work. In this work, we will treat the field theory as a phenomenological low energy effective theory. 
We consider the following Lagrangian density:
\begin{align}
\mathcal{L}_\text{LC}=
&\frac{Q_{IJ}}{2\pi} \epsilon^{\mu\nu\lambda\sigma}  b^I_{\mu\nu}\partial_\lambda a^J_\sigma
 + \frac{\Theta}{8\pi^2} R_{IJ} \epsilon^{\mu\nu\lambda\sigma}   \partial_\mu a^I_\nu \partial_\lambda a^J_\sigma \nonumber \\
& +j^I_\mu a^I_\mu + \mathcal{J}^I_{\mu\nu} b^I_{\mu\nu}
\label{FTlag}
\end{align}
with the summation of repeated indices implicitly assumed. Here $a^I_\mu$ for each $I$ is a 1-form $U(1)$ gauge field that couples to the current $j^I_\mu$ of non-trivial point excitations, and $b^I_{\mu\nu}$ for each $I$ is a 2-form $U(1)$ gauge field that couples to the 2-form current $\mathcal{J}^I_{\mu\nu}$ of non-trivial string excitations. ($b_{\mu\nu}^I$ and $\mathcal{J}_{\mu\nu}^I$ are antisymmetric in permutation of $\mu,\nu$.) Their gauge transformations are given by $a^I_\mu\rightarrow a^I_\mu+\partial_\mu \alpha^I$ and $b^I_{\mu\nu}\rightarrow b^I_{\mu\nu}+\partial_\mu \beta^I_\nu + \partial_\nu \beta^I_\mu$. $Q_{IJ}$ and $R_{IJ}$ are integer-valued non-singular matrices, and $R$ is symmetric. $\Theta$ is an ``order parameter field" which is extrinsically determined and does not have dynamics. When we consider the compactification radius of each $a_\mu^I$ as $2\pi$, the integral $\frac1{8\pi^2}\int d^4xR_{IJ} \epsilon^{\mu\nu\lambda\sigma}   \partial_\mu a^I_\nu \partial_\lambda a^J_\sigma$ is quantized to integer values for all closed space-time manifolds. Therefore the partition function of the system is invariant under the transformation $\Theta\rightarrow \Theta+2\pi$. For a manifold with boundary, $\Theta\rightarrow \Theta+2\pi$ will induce a boundary Chern-Simons term $\int_{\rm boundary}d^3n_\mu\frac{1}{4\pi}R_{IJ}\epsilon^{\mu\nu\sigma\tau}a_\nu^I\partial_\sigma a_\tau^J$. Here $\int dn_\mu$ is the boundary volume integration with $dn_\mu$ along the normal direction. 
The topological field theory (\ref{FTlag}) has the form of BF theory\cite{birmingham1991,hansson2004}. Similar theory has been proposed to describe fermionic and bosonic topological insulators\cite{cho2011,vishwanath2013}, and the Walker-Wang model\cite{walker2012,simon2013}. Compared with previous works, the essential new ingredient in our theory (\ref{FTlag}) is the string-string braiding statistics, which, as will be explained in the following, is enabled by the possibility of $\Theta$ vortex loops.

We start from a system with constant $\Theta$. The equation of motion is given by
\begin{align}
&j^I_\mu =-\frac{1}{2\pi}Q_{JI}\epsilon^{\mu\nu\lambda\sigma} \partial_\nu B^J_{\lambda\sigma}, \nonumber \\
&\mathcal{J}^I_{\mu\nu} =-\frac{1}{2\pi}Q_{IJ} \epsilon^{\mu\nu\lambda\sigma} \partial_\lambda a^J_\sigma.
\end{align}
Therefore, the string particle carries flux of $a_\mu^I$, so that braiding the $I^\text{th}$ particle $\Pi_I$ (whose current is given by $j^I_\mu$) and the $J^\text{th}$ string  $\Sigma_J$ (whose current is given by $\mathcal{J}^J_{\mu\nu}$) produces the Berry's phase
\begin{align}
\theta_{\Pi_I,\Sigma_J}=-2\pi (Q^{-1})_{IJ}.
\end{align}
By properly choosing the matrix $Q$, the Lagrangian density $\mathcal{L}_\text{CL}$ can capture the particle-string statistics in the layer constructed model. For example, to capture the braiding statistics in Eq. (\ref{EMstat}), we need to take $Q=p$ (where $Q$ is a $1\times 1$ matrix). For the case studied in Sec. \ref{Coexist}, we should have $Q=3$. For these two cases, $R$'s are taken to be $0$. However, even with non-zero matrix $R$, the string-string braiding is trivial when $\Theta$ is a constant. To obtain nontrivial string-string braiding, it is essential to consider space-time dependent $\Theta$.

\begin{figure}[t]
\centerline{
\includegraphics[width=3
in]{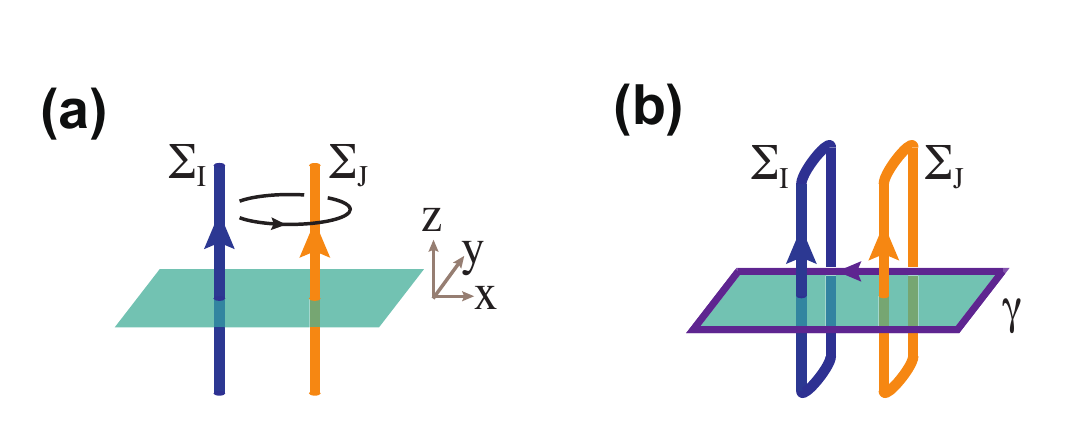}
}
\caption{\label{FT} (a) The configuration of two strings $\Sigma_I$ and $\Sigma_J$ that are aligned along the $z$ direction and penetrate the $z=0$ plane (green) where $\Theta$ jumps by $2\pi$. The strings $\Sigma_I$ and $\Sigma_J$ have non-trivial braiding. (b) The configuration of a finite open surface $\gamma$ across which $\Theta$ jumps by $2\pi$ together with strings $\Sigma_I$ and $\Sigma_J$ that linked with the vortex loop $\partial \gamma$ (purple).
}
\end{figure}

Now, we can consider the case with specially varying $\Theta$. The equation of motion is given by
\begin{align}
&j^I_\mu =-\frac{1}{2\pi}Q_{JI}\epsilon^{\mu\nu\lambda\sigma} \partial_\nu B^J_{\lambda\sigma}- \frac{1}{4\pi^2} R_{IJ} \epsilon^{\mu\nu\lambda\sigma} \partial_\nu\Theta \partial_\lambda a^J_\sigma , \nonumber \\
&\mathcal{J}^I_{\mu\nu} =-\frac{1}{2\pi}Q_{IJ} \epsilon^{\mu\nu\lambda\sigma} \partial_\lambda a^J_\sigma.
\end{align}
For the case without any point particles, i.e. $j^I_\mu=0$, we have
\begin{align}
(Q^{-1T} R Q^{-1})_{IJ} \mathcal{J}^J_{\mu\nu} \partial_\nu \Theta= \epsilon^{\mu\nu\lambda\sigma}\partial_\nu B^I_{\lambda\sigma}.
\end{align}
Since the bulk of system is invariant under the transformation $\Theta\rightarrow \Theta+2\pi$, we can consider a $2D$ surface in the system across which the the value of $\Theta$ jumps by $2\pi$. For example, when this $2D$ surface is the $xy$ plane, we have
\begin{align}
\partial_\mu \Theta= \delta_{\mu,z} 2\pi \delta(z).
\label{ThetaConfig}
\end{align}
When the vortex string $\Sigma_J$ goes through the $z=0$ plane, the point of penetration acts as a point source for $\epsilon^{\mu\nu\lambda\sigma}\partial_\nu B^I_{\lambda\sigma}$. If we consider braiding strings $\Sigma_I$ and $\Sigma_J$ as is shown in Fig. \ref{FT} (a) (with the $z=0$ plane where $\partial_z \Theta=2\pi$ the green plane shown in the figure), the Berry's phase of this braid is
\begin{align}
\omega_{\Sigma_I,\Sigma_J}=2\pi(Q^{-1T} R Q^{-1})_{IJ}.
\label{FTStrBrd}
\end{align}

In this discussion, the only non-trivial effects from the $\Theta$ configuration in Eq. (\ref{ThetaConfig}) is the braiding statistics between strings $\Sigma_I$ and $\Sigma_J$ that go through the $z=0$ only once and close at infinity. It is straightforward to see that if we consider strings of finite size, they have to cross the $z=0$ plane even number of times and their braiding is always trivial. Therefore, the system behaves the same as the original system without the jump of $\Theta$ across $z=0$. To obtain non-trivial braiding of finite-size strings, we need to consider 2D surface $\gamma$ with boundary $\partial \gamma \neq \emptyset$ crossing which $\Theta$ jumps by $2\pi$. Such a surface is a vortex loop of $\Theta$ at $\partial\gamma$. The strings that are linked with the vortex loop will have non-trivial string braiding statistics. For example, the braid statistics of the string $\Sigma_I$ and $\Sigma_J$ in Fig. \ref{FT} (b) is given by Eq. (\ref{FTStrBrd}). Compare to the result we present earlier in layer construction, we can easily identify the defected layer in Fig. \ref{TwoLoopsDis} (a) as the surface $\gamma$ and the edge dislocation as $\partial \gamma$. Since the jump of $\Theta$ is not locally observable in the bulk, we can think of the boundary $\partial \gamma$ as a line defect which is consistent with the edge dislocation interpretation.

For a specific layer constructed model, we can choose proper $Q_{IJ}$ and $R_{IJ}$ to capture the phenomenology of the model. For example, for the example we studied in Sec. \ref{GenericState}, the matrix $Q_{IJ}$ is determined by the statistics between point particles and string excitations in Eq. (\ref{pt_f-str}):
\begin{align}
Q=\left(
\begin{array}{cc}
p/2 & 0 \\ 0 & 2
\end{array}
\right).
\end{align}
With the choice
\begin{align}
R=\left(
\begin{array}{cc}
0 & 1 \\ 1 & 0
\end{array}
\right),
\end{align}
we obtain the string braiding statistics
\begin{align}
2\pi(Q^{-1T} R Q^{-1})_{IJ} =2\pi\left(
\begin{array}{cc}
0 & 1/p \\ 1/p & 0
\end{array}
\right),
\end{align}
which is consistent with braiding statistics $\omega^d_{c,f}$ between $c$-string and $f$-string calculated in Eq. (\ref{cfbraid}) in the presence of the $d$-string.

\section{General Discussion on String braiding}
\label{GeneralStrBrd}
\subsection{General Identities on Abelian String Braiding Statistics}
\label{GeneralA}

\begin{figure}[t]
\centerline{
\includegraphics[width=3.5
in]{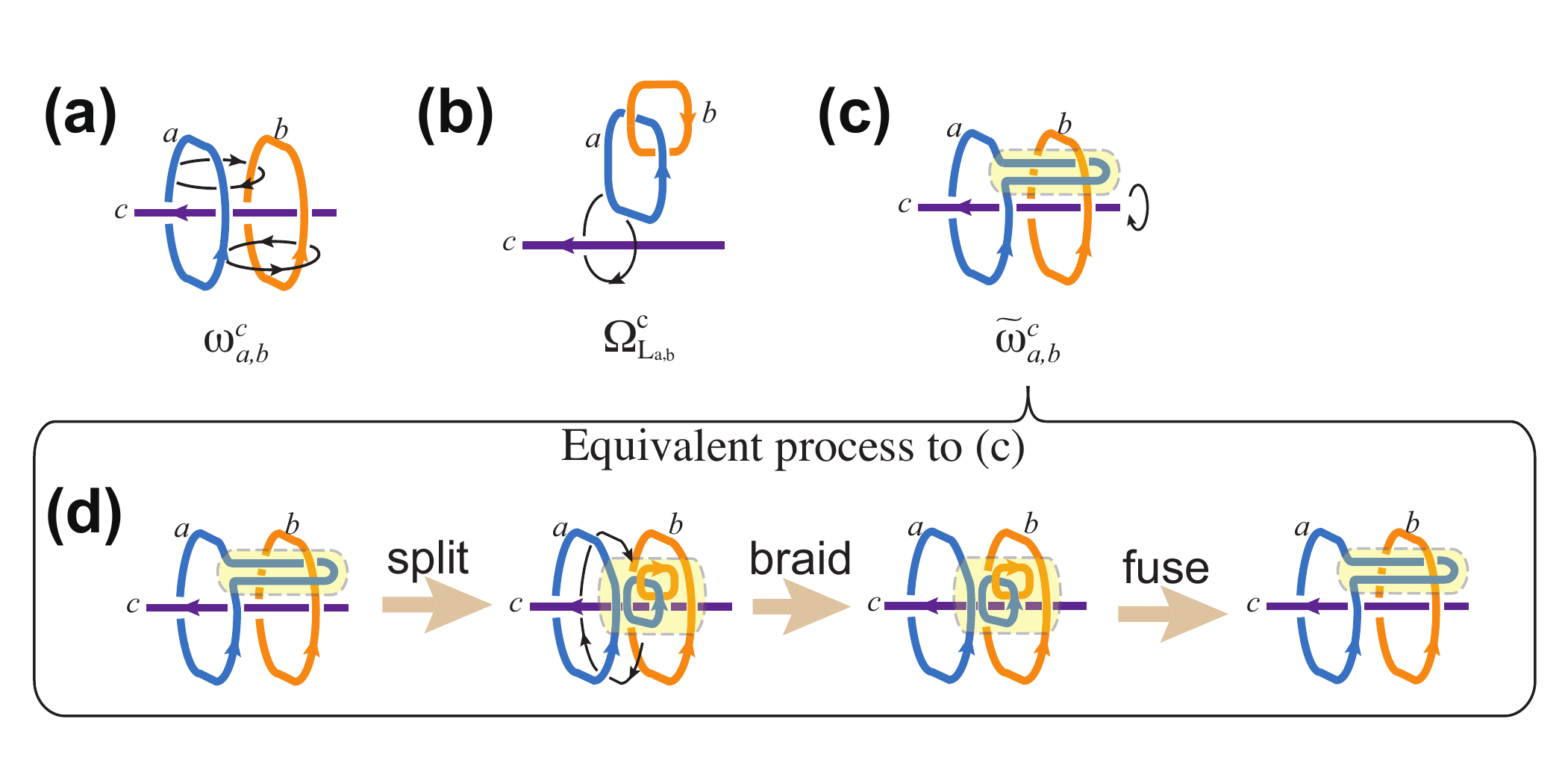}
}
\caption{\label{StringBrd} Three different string braiding processes, named as the string-string braiding (a), link-string braiding (b) and the linked braiding (c). As is explained in the main text, the Berry's phase associated to these three processes are identical. (d) The decomposed steps that relate processes in (c) and (b).
}
\end{figure}

In the last part of previous section, we discuss the braiding statistics of strings that involves two dynamical string excitations and one extrinsic string defect. It is conceivable that similar string braiding can happen in the same configuration for three dynamical string excitations. As is recently discussed in Ref. \onlinecite{Levin2014StrBrd,Ran2014StrBrd,JuvenWenStrBrd2014}, in twisted lattice gauge theories defined by non-trivial group cohomology classes, different flux strings can braid non-trivially. Ref. \onlinecite{Levin2014StrBrd} proposed an identity on braiding statistics of these flux strings. For two strings $a$ and $b$ both linked with $c$ (Fig. \ref{StringBrd} (a)), the braiding phase of $a$ and $b$ is defined as $\omega_{a,b}^c$. Ref. \onlinecite{Levin2014StrBrd} proved the following identity (up to $2\pi$ times integers) for 3D $Z_N^k$ gauge theories:
\begin{eqnarray}
N(\omega_{a,b}^c+\omega_{b,c}^a+\omega_{c,a}^b)=0.\label{TriId2ml}
\end{eqnarray}
In the following, we will present a more general proof of a slightly modified version of this identity for Abelian string braiding, by making use of splitting and fusion of strings.

As a general setup of our discussion, we will consider the string excitations that has the following properties: (1) The strings are Abelian in sense that there are no non-trivial degenerate states associated to a local string configuration; (2) The strings have no point-like excitation attached to them. (3)  Strings of the same type can split and fuse without inducing additional phase factor; (4) The strings can link with each other. 

Since we have assumed that there is no non-trivial degeneracy associated with the string configuration, the Berry's phase $\omega_{a,b}^c$ is simply a $U(1)$ phase. This Berry's phase $\omega_{a,b}^c$ should be a topological invariant that does not depend on the shape of the strings in the process of braiding. In Sec. \ref{GenericState}, we have shown that by fusion and splitting of the same type of strings, one can topologically deform the string-string braiding process of $a$, $b$ strings linked to $c$ (Fig. \ref{StringBrd} (a)) to a particle-string braiding process in which the ``particle" $L_{a,b}$ is formed by two linked loops of types $a$ and $b$. Denoting the Berry's phase obtained by the braiding of $L_{ab}$ with string $c$ as $\Omega_{L_{a,b}}^c$, we obtain the equation
 $\omega_{a,b}^c=\Omega_{L_{a,b}}^c$. Now we consider a different three-string configuration shown in Fig. \ref{StringBrd} (c). In this configuration, every pair of strings are mutually linked, with linking number 1. The ``braiding" process (which will be referred to as the linked braiding) in this configuration is defined to be a $2\pi$ rotation the two string $a$ and $b$ about the string $c$ (see Fig. \ref{StringBrd} (c)) while keeping the relative position between $a$ and $b$ fixed. The Berry's phase associated with this process is denoted as $\tilde{\omega}^c_{a,b}$. Following similar string fusion and splitting procedure as has been presented earlier in Fig. \ref{StringBrd} (c), we can deform the linked braiding process to the same particle-string braiding shown in Fig. \ref{StringBrd} (b). This is illustrated in detail in Fig. \ref{StringBrd} (d).
 Therefore, we conclude,
\begin{align}
\omega_{a,b}^c=\Omega_{L_{a,b}}^c =\tilde{\omega}^c_{a,b}.
\label{BraidEquiv2}
\end{align}

\begin{figure}[t]
\centerline{
\includegraphics[width=1.5
in]{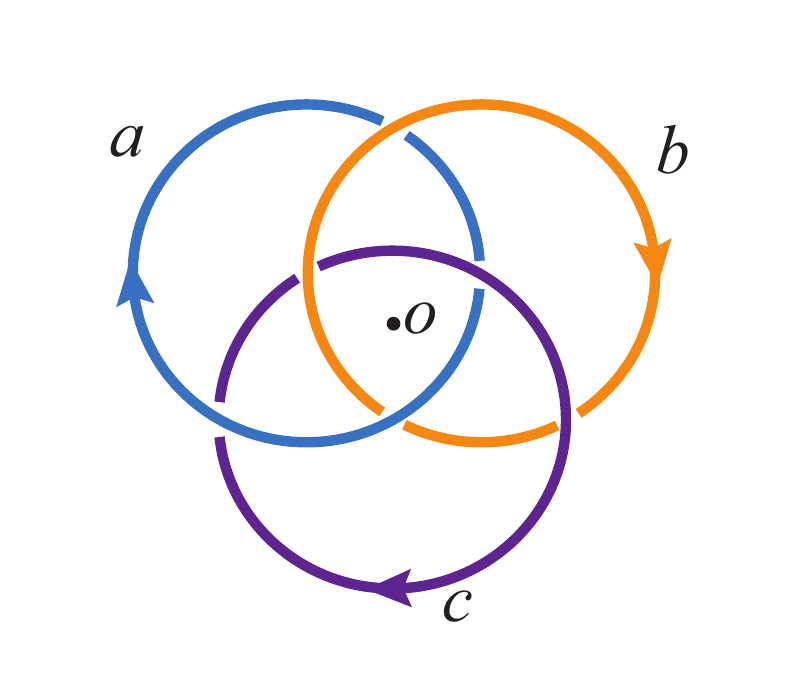}
}
\caption{\label{ThreeLoop} The configuration of three mutually linked strings. This configuration is topologically equivalent to the configuration shown in Fig. \ref{StringBrd} (c), but is deformed to a shape with three-fold rotation symmetry according to an axis perpendicular to the plane at the point $o$.
}
\end{figure}

Now we proceed to prove the a modified version of the identity (\ref{TriId2ml}) by proving it for $\tilde{\omega}^c_{a,b}$. The configuration of three mutually linked loops in the Fig. \ref{StringBrd} (c) can be topologically deformed to that in Fig. \ref{ThreeLoop}. This configuration is symmetric upon cyclic permutation of the three strings. The ``linked braiding" process defined in Fig. \ref{StringBrd} (c) can be rephrased as the operation that keeps the string $c$ fixed and rotates simultaneously the strings $a$ and $b$ about an axis, which goes through the reference point $o$ and is perpendicular to the 2D plane, clock-wisely by $2\pi$. This process is the linked braiding of the string $a$ and $b$ with respect to the string $c$. Due to the cyclic symmetry between the three strings in this configuration in Fig. \ref{StringBrd}, we can also consider similar linked braiding processes between the strings $b$ and $c$ with respect to the string $a$, and that between $c$ and $a$ with respect to $b$. If we perform these three different types of linked braiding sequentially, the net Berry's phase obtained is by definition $\tilde{\omega}_{a,b}^c+\tilde{\omega}_{b,c}^a+\tilde{\omega}_{c,a}^b$. However, one can easily see that the combined process is actually a global $4\pi$ rotation of the whole string configuration around axis $o$, since each string is rotated by $2\pi$ during two of the three processes, and is held static during one of them.
Since a $4\pi$ rotation is a topologically trivial action in three dimension, the total Berry's phase induced by the rotation must be trivial. This implies that the sum of three different linked braiding Berry's phases should be trivial, namely
\begin{align}
\tilde{\omega}_{a,b}^c+\tilde{\omega}_{b,c}^a+\tilde{\omega}_{c,a}^b=0.
\label{TriId1}
\end{align}
Therefore by using Eq. (\ref{BraidEquiv2}), we prove the following identities:
\begin{align}
&\omega_{a,b}^c+\omega_{b,c}^a+\omega_{c,a}^b=0,
\label{TriId3} \\
&\Omega_{L_{a,b}}^c+\Omega_{L_{b,c}}^a+\Omega_{L_{c,a}}^b=0.
\label{TriId4}
\end{align}
Eq. (\ref{TriId3}) directly implies the previously proposed Eq. (\ref{TriId2ml}). Eq. (\ref{TriId4}) constraints the braiding statistics of strings and ``link particles".

Our proof here only relies on adiabatical deformations and fusion-splitting processes of strings, which is thus applicable to generic 3D topologically ordered states that involves Abelian string-like excitations. 

\subsection{Non-Abelian String Braiding Statistics}
\label{GeneralB}

The discussion above only applies to the 3D topological states with string excitations that satisfies the assumptions given in the beginning of the previous subsection. It is worthy to point out that there are certainly possibility of other more generic string like excitations that violates some of these assumptions. For instance, the same string configurations can in principle carry non-Abelian degrees of freedom. The general framework of non-Abelian string-string braiding is beyond the scope of this paper, but we would like to present one explicit example of a system with non-Abelian string-string braiding.

Consider two Weyl fermions in $(3+1)$-dimensions with opposite chirality, which are equivalent to a Dirac fermion. Each Weyl fermion is time-reversal invariant, and a superconducting pairing can be introduced within each Weyl fermion. The Fermi level can be either at or away from the Weyl point, which does not affect our discussion. The BCS Hamiltonian of such a superconductor can be written as
\begin{align}
\hat{H}=\sum_{\bf{k}}  \left[ \psi^\dag_{R\textbf{k}} (v_F\sigma\cdot \textbf{k}-\mu) \psi_{R \bf{k}}
 +\psi^\dag_{L\bf{k}} (-v_F\sigma\cdot \textbf{k}-\mu) \psi_{L\bf{k}} \right]+ \nonumber \\
\frac{1}{2}\sum_{\bf{k}} \left[
|\Delta_R|e^{i\theta_R}\psi^\dag_{R\textbf{-k}} i\sigma_y \psi^\dag_{R \bf{k}}
+|\Delta_L|e^{i\theta_L}\psi^\dag_{L\textbf{-k}} i\sigma_y \psi^\dag_{L \bf{k}}+h.c.
\right],\label{TSCHamiltonian}
\end{align}
where $\psi_{R \bf{k}}$ and $\psi_{L \bf{k}} $ are two-component fermion operators, $v_F$ is the fermi velocity, $\sigma$'s are the $2\times2$ Pauli matrices, and $\Delta_R=|\Delta_R|e^{i\theta_R}$ and $\Delta_L=|\Delta_L|e^{i\theta_L}$ are the pairing order parameters on the fermi surface of $\psi_{R \bf{k}}$ and $\psi_{L \bf{k}}$. Without the pairing term and with $\mu>0$, the Fermi surface of $\psi_{R \bf{k}}$ carries Chern number 1 while the Fermi surface of $\psi_{L \bf{k}}$ carries Chern number -1. This model was studied before\cite{Qi2013TSC} as a minimal model of time-reversal invariant topological superconductivity (TSC)\cite{QiTSC2009,Roy2008,Schnyder2008}, which describes a TSC when $\theta_L=0,~\theta_R=\pi$, and a trivial superconductor when $\theta_L=\theta_R=0$\cite{qi2010b}.

\begin{figure}[t]
\centerline{
\includegraphics[width=2.5
in]{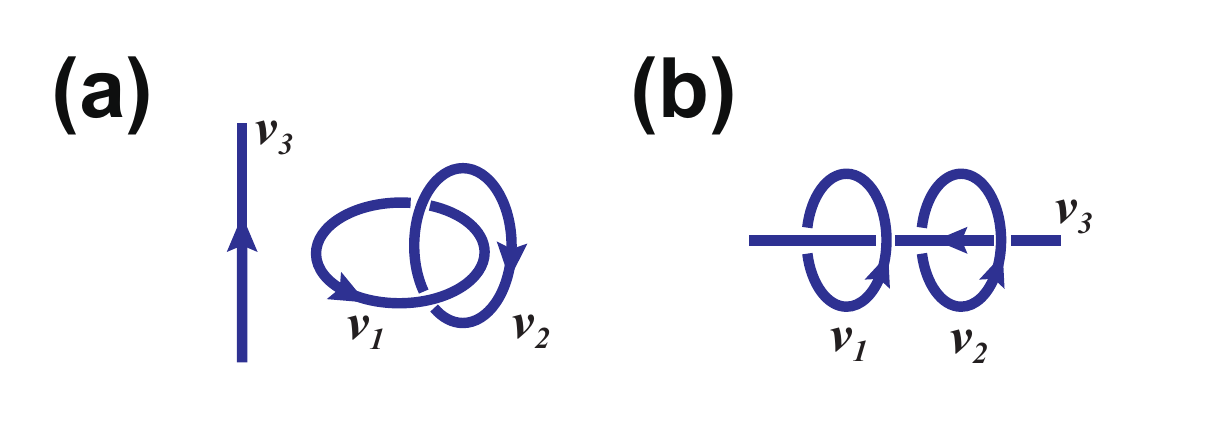}
}
\caption{\label{TSC} (a) and (b) show two different configurations of chiral vortex strings. 
}
\end{figure}

As a probe to the TSC, Ref. \onlinecite{Qi2013TSC} studied the chiral vortex strings in this system, which are $2\pi$ vortex strings of only one of the pairing phases, such as $\theta_R$.
This chiral vortex string carries a $(1+1)$-d chiral Majorana fermion mode, which has a axial anomaly. For the purpose of this paper, we will focus on the zero energy mode in the spectrum of the chiral Majorana fermion, usually named as the Majorana zero mode\cite{kitaev2001}. A contractible vortex string does not carry a zero energy mode, since the Majorana fermion has anti-periodic boundary condition. (Without going into a detailed calculation, one can see that this must be true since a Majorana zero mode, if exists, cannot disappear when the loop shrinks to a point and disappear. Therefore it cannot occur in a contractible loop.) When a vortex string $v_1$ is linked with another chiral vortex string $v_2$ (see Fig. \ref{TSC} (a)), there will be a Majorana zero modes $\gamma_{1(2)}$ on each vortex string. Each Majorana zero mode is separated from other chiral modes by a finite size gap that is inversely proportional to the length of the vortex strings. There are two degenerate states with the opposite fermion parity $(-1)^F\equiv i \gamma_1 \gamma_2$ associated with the two zero modes on the two linked strings. If we braiding the link between $v_1$ and $v_2$ around a third chiral vortex string $v_3$, the Berry's phase is equal to
$(-1)^F$. Therefore, we see that the string braiding statistics relies on the state in the degenerate space defined by the string configuration, which is a three-dimensional generalization of ``fusion channels" of a non-Abelian anyon. Consider a different chiral vortex string configuration in Fig. \ref{TSC} (b) with two vortex strings $v_1$ and $v_2$ both linked with $v_3$. In this configuration both $v_1$ and $v_2$ carry Majorana zero modes $\gamma'_1$ and $\gamma'_2$, while $v_3$ does not have a zero mode. If we braid the vortex string $v_1$ and $v_2$ in the way defined in Fig. \ref{TwoLoopsDis} (a), the Majorana fermion operators $\gamma'_{1,2}$ will each obtain a $-1$ sign after a full braiding, due to the phase winding around the vortices. Therefore one expects that the braiding between these two strings with zero modes is the same as the braiding of two point-like vortices in $2+1$-d $p+ip$ superconductor\cite{read2000,ivanov2001}, up to an undetermined overall phase factor. The non-Abelian Berry's phase is a unitary operator acting on the two-dimensional Hilbert space, which can be written as $e^{i\frac{2\pi i \gamma'_1 \gamma'_2}{4}}$.
Since the chiral vortices are generically confined with the anti-chiral vortices energetically\cite{Qi2013TSC}, they have to be considered as external defects and the Abelian phase factor of the braiding process is not well-defined.

Since we can consider all the chiral vortices to be identical, we can also consider the ``half braid" between $v_1$ and $v_2$ that is similar to the string braiding defined Fig. \ref{TwoLoopsDis} (a) but only exchanges the position of two vortex strings instead of braiding one string around the other. In this half braid process, we can keep the sizes of all three vortex strings $v_1$, $v_2$ and $v_3$ finite. The degenerate space defined by the two zero modes $\gamma'_{1,2}$ are separated from other excited states by a finite size gap. Therefore the half braid should not change the total fermion parity $(-1)^F=i \gamma'_1 \gamma'_2$ of the three-string configuration. The only possible non-trivial operation to the Majorana zero modes is
\begin{align}
&\mathcal{U}^\dag \gamma'_1 \mathcal{U} = \gamma'_2, \nonumber \\
&\mathcal{U}^\dag \gamma'_2 \mathcal{U} = -\gamma'_1,
\end{align}
where $\mathcal{U}$ is the unitary transformation induced by the half braid. This determines $\mathcal{U}=e^{\frac{2\pi i \gamma'_1 \gamma'_2}{8}}$ up to a $U(1)$ phase.


When more generic configurations of vortex strings are considered, one can see more qualitative difference from the Abelian case. For example, one can consider the configuration of three mutually linked chiral vortex strings shown in Fig. \ref{ThreeLoop}. In the Abelian case, a ``linked braiding" procedure is defined in this configuration (Fig. \ref{StringBrd} (c) and (d)), which is proved to induce the same Berry's phase as the three-string braiding process in Fig. \ref{StringBrd} (a). In the case of chiral vortex strings, the configuration of three mutually linked strings is very different from two strings linked with the third one: When each loop is linked with two other loops, there is no Majorana zero mode on any of the three loops. Consequently there cannot be any nontrivial non-Abelian Berry's phase in the linked braiding process. If the linked braiding and unlinked braiding (the procedures shown in Fig. \ref{StringBrd} (a) and (c)) are still related, the linked braiding will at most be related to a given fusion channel of the two strings in the unlinked configuration.
In general, the fusion and splitting of non-Abelian strings involve the change of topological degeneracy, which is more complicated than the case of Abelian strings with ``adiabatic" fusion and splitting within the same species. Fusion channels have to be defined when we consider the splitting and fusion of strings. We will leave this for future study.

One can consider a multi-component generalization of the model (\ref{TSCHamiltonian}), which has a matrix-valued pairing order parameter, and allows monopole-like non-Abelian point defects in addition to the chiral vortices\cite{teo2010,ran2011,freedman2011,freedman2011b}. The non-Abelian statistics of monopole defects are also related to Majorana zero modes, and it is interesting to investigate the interplay of the braiding statistics of monopoles and chiral vortex strings.

There are also other similar examples in the 3D quantum double model\cite{Wen2014StrBrd} and 3D Dijkgraaf-Witten lattice theory\cite{JuvenWenStrBrd2014} that strings can carry non-Abelian degrees of freedom and have non-Abelian braiding statistics. When we dimensionally reduced the $z$ direction to a finite compactified circle for these lattice gauge theory systems, the strings align along the $z$ direction which can be viewed as a quasi-particle in the 2D sense can be asigned quantum dimensions. Since the 3D lattice guage theory, when dimensionally reduced, becomes 2D lattice guage theory. Each quasi-particles (in the 2D theory) must have integer quantum dimension, and so are the strings (along the $z$ direction). In constrast, if we dimensionally reduce the 3D TSC to 2D with a $\pi$ chiral flux in the $z$ direction cycle, the chiral vortex strings (along the $z$ direction) will have quantum dimension $\sqrt{2}$ due to the Majorana zero mode. It would be interesting to find more examples of string excitations with non-integer quantum dimensions. Moreover, Ref. \onlinecite{Levin2014StrBrd,JuvenWenStrBrd2014,Ran2014StrBrd} mainly concerns the braiding process defined in Fig. \ref{StringBrd}  (a) between flux strings in the twisted Dijkgraaf-Witten lattice gauge theory. It would also be interesting to study the braiding process defined in Fig. \ref{StringBrd} (b) and (c) for these models.


\section{Conclusion and Discussion}
\label{conclusion}
In this paper, we propose a general approach to obtain non-trivial 3D topological states through layer construction. Starting from stacked layers of 2D Abelian topological states, anyon-condensation induces inter-layer coupling and drives the system into a three-dimensional state which may have interesting bulk and/or surface topological orders. We illustrate several different possibilities with explicit examples. The first class of states are those with trivial bulk topological order and nontrivial surface two-dimensional topological order. We obtain general criteria for such states, and show that this state has the same topological order on surfaces with different orientation, even though it appears to be anisotropic.
This implies that the layer construction indeed captures the properties of homogeneous 3D states. The second class of states are those with nontrivial bulk topological order, which means nontrivial ground state degeneracy on a three-dimensional torus. The simplest examples realize the discrete gauge theories. More generically, the layer construction can describe broader class of topologically ordered states which have
 coexisting surface and bulk topological order. One of the most interesting properties that can be realized in some layer-constructed topological states is the string-string braiding statistics, in addition to the more conventional particle-string braiding that exists in discrete gauge theories. We provide an explicit example of a state with non-trivial string-string braiding, and also discuss the relation between string-string braiding and particle-string braiding by making use of fusion and splitting procedures of strings. A topological field theory is proposed
 to describe the topological order and string braiding statistics of the layer constructed systems. The topological field theory has the form of BF theory ``twisted" by an ``axionic" topological coupling of the gauge field with an axion field $\theta$. Based on the explicit examples and field theory, we further discuss the general properties of three-string braiding, and give a more general proof of the string braiding identities Eq. (\ref{TriId1}), (\ref{TriId3}) and (\ref{TriId4}) with Eq. (\ref{TriId3}) being a a modified version of the previously proposed identity (\ref{TriId2ml}) proved for $Z_N^k$ lattice gauge theories. Finally, we discussed the generalization of the 3D topological order and string-string braiding to non-Abelian systems by proposing one example system which is a superconductor with chiral vortex strings. Compared with Abelian strings, the non-Abelian strings have multiple fusion channels, non-Abelian braiding phase, and topological degeneracy depending on string linking.

There are a lot of interesting open questions along the direction of this work. The first question is how to write down lattice models which realizes the coupled layers and anyon condensation. 
For layers of $Z_p$ toric code with condensation described in Fig. \ref{LGT3d} (a), the lattice model is simply the 3D toric code. This can be understood in the following way. Take $p=2$ as an example. Consider the cubic lattice with $Z_2$ degrees of freedoms defined on each link. The vertex term on the vertex $v$ is given by $\mathcal{O}_v=-\prod_{l \text{ connects to v} } \sigma^z_l$ with $l$ labeling the link, and the plaquette term at plaquette $p$ is given by $\mathcal{O}_p=-\prod_{l \in \text{ plaquette }p } \sigma^x_l$. Now consider adding a term $h \sigma^z_l$ for all vertical links with $h$ a coupling constant. In the $h\rightarrow \infty$ limit, all the vertical links are polarized, 3D toric code splits up into layers of 2D toric code. As we lower the coupling $h$, a transition that is described in Fig. \ref{LGT3d} (a) will happen and bring the system back to the topological ground state of 3D toric code. Its $Z_p$ and twisted version can also be constructed in a similar fashion. It would be interesting to construct lattice models that capture the physics of more generic 3D states obtained through layer construction.

Another interesting question is the relation of the layer constructed states with the Walker-Wang models. Some similarities are discussed in Sec. \ref{Coexist} between these two systems, and the Walker-Wang's work\cite{walker2012} set a general mathematical framework of defining 3D topological states. However, string-string braiding has not been discussed in Walker-Wang model, so it will be an interesting future direction. 

 If incorporated with symmetry, the layer construction for 3D state with purely surface topological order can be used to build models for symmetry protected topological (SPT) states with gapped surfaces, as has been discussed in Ref. \onlinecite{wang2013}. Our analysis on the side surface can serve as a Hamiltonian formalism for these topologically ordered surfaces of 3D SPTs which allow us to study more carefully the symmetry breaking transition on the surface.

Another future direction is a more systematic understanding of the 3D topological order. For 3D gapped state with purely surface topological order and trivial bulk topological order, we obtained some general criteria on when it occurs and what is the surface topological order (Sec. \ref{GCpSTO}). However, for more generic 3D states with bulk and/or surface topological order, we have only done a case-by-case study of the topological order for each given set of condensed anyons. More specifically, what is needed is a general formula which determines the topological properties of the 3D state (such as the $Q$ and $R$ matrices in the field theory approach) from the ``microscopic" data, {\it i.e.} the layer $K$ matrix before the anyon condensation, and the condensed anyons given by $p_i$ and $q_i$ vectors. It is also interesting to consider anyon condensation and layer construction in non-Abelian layers, using the technique developed in literature\cite{Bais2009, Bais2013,Kong2013}. A specific question will be how to obtain the topological superconductor state with chiral vortex from layer construction.


Moreover, we would like to make some more general comments on the string braiding statistics in 3D topologically ordered states. In Sec. \ref{GeneralStrBrd}, we have shown that, for Abelian strings with trivial fusion and splitting within the same species, the string braiding statistics can be defined in various of different ways with configurations of different linking numbers, which yield exactly the same Berry's phase. This implies that linked string configuration (Fig. \ref{ThreeLoop}) might the objects that play a central role in 3D topological order in general. For example, by applying our analysis on the identification between three-string braiding (Fig. \ref{StringBrd} (a)) and the link-string braiding (Fig. \ref{StringBrd} (b)) to the  Dijkgraaf-Witten lattice $Z_N^k$ or $Z_{N_1}\times Z_{N_2}\times Z_{N_3}\times Z_{N_4}$ gauge theories studied in Ref. \onlinecite{Levin2014StrBrd,Ran2014StrBrd, JuvenWenStrBrd2014}, we immediately conclude that the link between two flux strings is in fact charged under the gauge group. Further, our example of chiral vortex in TSC and the result in Ref. \onlinecite{Wen2014StrBrd, JuvenWenStrBrd2014} show that the link of flux strings or themselves can even carry non-Abelian degrees of freedom. It would be very interesting to construct the mathematical framework that captures the braiding, link and fusion of the strings and study the various operations to the linked string configurations discussed in Sec. \ref{GeneralStrBrd} in generality.

At a final comment, we would like to memtion that Ref. \onlinecite{Kong2014BF} recently proposed a general theory for topological order in any dimension. It would be interesting to study its connection to the layer construced models and to the general properties of string statistics in 3D.

{\it Acknowledgements} We would like to thank Maissam Barkeshli, Ling-Yan Hung, Senthil Todadri, Chong Wang, Zhenghan Wang and Xiao-Gang Wen for help discussions. This work is supported by the BOCO fellowship (CMJ) and David and Lucile Packard Foundation (XLQ).

\bibliography{TI}

\end{document}